\documentclass[10pt]{article}
\ifx\pdfoutput\undefined
\usepackage[dvips]{graphicx}
\usepackage{psfrag}
\else
\usepackage[pdftex]{graphicx}
\usepackage{type1cm}
\fi 
\usepackage{hyperref}
\def\stackunder#1#2{\mathrel{\mathop{#2}\limits_{#1}}}%
\def\eqa{\begin{eqnarray}}
\def\eqae{\end{eqnarray}}

\def\stackunder#1#2{\mathrel{\mathop{#2}\limits_{#1}}}%

\def\@magscale#1{ scaled \magstep #1}

\catcode`@=11  
\def\un#1{\relax\ifmmode\@@underline#1\else
        $\@@underline{\hbox{#1}}$\relax\fi}
\catcode`@=12

\def\b{\beta}

\def\l{\lambda}
\def\m{\mu}

\def\r{\rho}

\def\t{\tau}

\def\L{\Lambda}

\def\X{\Xi}

\def\dslash{\not{\hbox{\kern-2pt $\partial$}}}
\def\Dslash{\not{\hbox{\kern-4pt $D$}}}
\def\pslash{\not{\hbox{\kern-2.3pt $p$}}}
 \newtoks\slashfraction
 \slashfraction={.13}
 \def\slash#1{\setbox0\hbox{$ #1 $}
 \setbox0\hbox to \the\slashfraction\wd0{\hss \box0}/\box0 }
 
\def\plpl{\raise-2pt\hbox{$\raise3pt\hbox{$_+$}\hskip-6.67pt\raise0.0pt
\hbox{$^+$}\hskip 0.01pt$}}
\def\mimi{\raise-2pt\hbox{$\raise3pt\hbox{$_-$}\hskip-6.67pt\raise0.0pt
\hbox{$^-$}\hskip 0.01pt$}} 

\def\bo{{\raise.15ex\hbox{\large$\Box$}}}               
\def\TH{{\raise.2ex\hbox{$\displaystyle \bigodot$}\mskip-4.7mu \llap H
\;}}
\def\face{{\raise.2ex\hbox{$\displaystyle \bigodot$}\mskip-2.2mu \llap
{$\ddot\smile$}}}                                      

\def\leftrightarrowfill{$\mathsurround=0pt \mathord\leftarrow \mkern-6mu
        \cleaders\hbox{$\mkern-2mu \mathord- \mkern-2mu$}\hfill
        \mkern-6mu \mathord\rightarrow$}
\def\dvec#1{\vbox{\ialign{##\crcr
        \leftrightarrowfill\crcr\noalign{\kern-1pt\nointerlineskip}
        $\hfil\displaystyle{#1}\hfil$\crcr}}}           

\def\frac#1#2{{\textstyle{#1\over\vphantom2\smash{\raise.20ex
        \hbox{$\scriptstyle{#2}$}}}}}                   
\def\sfrac#1#2{{\vphantom1\smash{\lower.5ex\hbox{\small$#1$}}\over
        \vphantom1\smash{\raise.4ex\hbox{\small$#2$}}}} 
\def\bfrac#1#2{{\vphantom1\smash{\lower.5ex\hbox{$#1$}}\over
        \vphantom1\smash{\raise.3ex\hbox{$#2$}}}}       
\def\afrac#1#2{{\vphantom1\smash{\lower.5ex\hbox{$#1$}}\over#2}}    

\newskip\humongous \humongous=0pt plus 1000pt minus 1000pt

\newif\ifdtup

\begin{document}

\title{General Coordinate Transformations as the Origins of Dark Energy}
\author{ Vincent G.J. Rodgers\footnote{
vincent-rodgers@uiowa.edu} and Takeshi Yasuda\footnote{takeshi-yasuda@uiowa.edu}\\Department of Physics and Astronomy\\The University of Iowa\\Iowa City, IA  52242 USA
}

\maketitle

\begin{abstract}
In this note we demonstrate that the algebra associated with coordinate transformations might contain the origins of a scalar field that can behave as an inflaton and/or a source for dark energy.  We will call this particular scalar field the {\em diffeomorphism} scalar field. In one dimension, the algebra of coordinate transformations is the Virasoro algebra while the algebra of gauge transformations is the Kac-Moody algebra.  An interesting representation of these algebras corresponds to certain field theories that have meaning in any dimension.  In particular the so called Kac-Moody sector corresponds to Yang-Mills theories and the Virasoro sector corresponds to the diffeomorphism field theory that contains the scalar field and a rank-two symmetric, traceless tensor.  We will focus on the contributions of the diffeomorphism scalar field to cosmology.   We show that this scalar field can, qualitatively, act as a phantom dark energy, an inflaton, a dark matter source, and the cosmological constant $\Lambda $. 

\end{abstract}

\newpage
\section{Introduction}
Recently experimental cosmology has brought new challenges to the theoretical understanding of the universe. Until now, the standard hot big-bang cosmology has been very successful in explaining four important observed facts: (1) the expansion of
Universe, (2) the origin of the cosmic background radiation, (3) the
nucleosynthesis of the light elements, (4) the formation of galaxies and the
large scale structure of Universe. However, recent discoveries from Supernova
Cosmology Project [\ref{SNCP}] and WMAP [\ref{WM}] strongly suggest that the
standard cosmology needs to be revised. This new standard cosmology must now
explain%
\begin{eqnarray*}
&&\ \circ \ \ \mbox{Flat and accelerating Universe} \\
&&\ \circ \ \ \mbox{Rapid expansion of Universe in early period (Inflation)%
} \\
&&\ \circ \ \ \mbox{Composition of Universe: 73\% dark energy; 23\% dark
matter; 4\% baryons\thinspace }
\end{eqnarray*}%
The theoretical search for this new standard cosmology is very active and there is an
overwhelming amount of literature published where most of them
have the same theme: scalar fields.  In this work we ask if there is a guiding principle that might account for the presence of new fields in cosmology.  We posit that there is a symmetry that can accompany general relativity in providing an explanation for some of the new features found in cosmology.  This symmetry is not new and is already exploited in general relativity through general coordinate covariance.  However we will realize the symmetry using an approach used in string theory.  Since string theories are two dimensional field theories, direct contributions to gravity coming from curvature and the Einstein-Hilbert action are diminished because of the low dimension.  In fact in two dimensions the Einstein-Hilbert action can at most distinguish different topological structures.  Our starting principle will be the algebra of Lie derivatives which exists in any dimension including one dimension.  We will show that this algebra admits a field theoretic representation analogous to the  Yang-Mills gauge potentials that arise from the algebra of gauge transformations.

In one dimension, this field theory comes from the {\em coadjoint representation} [\ref{Ki1},\ref{Ki2},\ref{Wi2},\ref{DvNR}] of the one dimensional algebra of Lie derivatives called the Virasoro algebra. In string theory this representation is used to understand the geometric origin of gravitational anomalies.  As we will review here, one can understand the origins of two dimensional gauge and gravitational anomalies through the {\em Geometric Action} [\ref{AS},\ref{W1},\ref{LR1},\ref{LR2},\ref{RR}] as well as establishes the Lagrangian for the fields living in the coadjoint representation called the {\em Transverse Action} [\ref{BLR},\ref{BRY}].  For this work, the importance feature of the Transverse Action as compared to the Geometric Action is that the Transverse Action can exist in any dimension.  We will review the construction of the Transverse Action with both the algebra of gauge and coordinate transformations. 
We show that there is a naturally occurring scalar fields called the {\em diffeomorphism scalar field} (diff scalar) that arises naturally from the algebra of associated with coordinate transformations.   

As a matter of completeness we will keep this paper somewhat self contained and outline of this paper as follows.  We will first review the
Robertson-Walker cosmology and as well as four of the more popular scalar field theories discussed in the literature.  This will introduce our notation to the reader and put this work in a more familiar setting.  In the next section we discuss the method we use at arriving at the Lagrangian for the diff scalar.  
We are then able to discuss the cosmological implications of the diffeomorphism scalar field by combining the action for the diff scalar with the Einstein-Hilbert action in the next section.   
We show that the diff scalar field can, qualitatively, act as  phantom dark energy, an inflaton, a dark matter source, and the cosmological constant $\Lambda $.

\section{Standard Cosmology Review}

The standard cosmology is based on two postulates that Universe is
homogeneous and isotropic at large scales [\ref{Pe},\ref{We1},\ref{D1}]. These postulates are strongly
supported by the cosmic microwave background (CMB) radiation coming from
various regions of sky: the temperature of the CMB radiation from different
parts of sky are almost identical. These postulates are called the \textit{%
cosmological principle}. With proper choice of coordinates these
symmetry postulates thus allow us to write the metric in the form of
Robertson-Walker metric:%
\begin{equation}
ds^{2}=-dt^{2}+a^{2}\left( t\right) \left\{ {dr^{2}\over 1-kr^{2}}
+r^{2}\left( d\theta ^{2}+\sin ^{2}\theta \mbox{ }d\phi ^{2}\right) \right\}
\end{equation}%
Here we used the convention: $\hbar =c=1$. The function $a\left( t\right) $
is called the scale factor of the universe (\textit{cosmic scale factor}).
Rescaling this factor appropriately, the constant $k$ can be always taken to
be $0$, $+1$, or $-1$:%

\begin{eqnarray*}
&&\ k=+1:\mbox{ closed universe with positive spatial curvature}  \\
&&\ k=0:\mbox{ flat universe with zero spatial curvature}  \\
&&\ k=-1:\mbox{ open universe with negative spatial curvature}
\end{eqnarray*}%

Since current observations seems to prefers a flat universe, we will adopt $k=0$ throughout this paper. The cosmological principle also demands that all cosmic
tensors are maximally form-invariant with respect to the spatial
coordinates. As a consequence of this constraint the most general
form-invariant tensor for the energy-momentum tensor is given by:%
\begin{equation}
T_{00}=\rho \left( t\right) ,\ \ \ T_{0i}=0,\ \ \ T_{ij}=p\left( t\right)
g_{ij},\ \ \ \ \ \ i,j=1,2,3
\end{equation}%
where $\rho \left( t\right) $ and $p\left( t\right) $ are arbitrary
functions that depend only on time. Rewriting the energy-momentum tensor as%
\begin{equation}
T_{\mu \nu }=\left( \rho +p\right) U_{\mu }U_{\nu }+pg_{\mu \nu }
\end{equation}%
where $U_{\mu }$ is a ``four-velocity vector"%
\begin{equation}
U^{0}=1,\mbox{ \thinspace \thinspace \thinspace \thinspace \thinspace
\thinspace \thinspace \thinspace \thinspace \thinspace \thinspace }U^{i}=0%
\mbox{ \thinspace }\left( i=1,2,3\right)
\end{equation}%
we identify $\rho \left( t\right) $ and $p\left( t\right) $ as with energy
density and pressure of a perfect fluid. 

Using this form of the energy-momentum tensor and the Robertson-Walker
metric, we obtain the \textit{Friedmann equations} from Einstein's field
equations:

\begin{eqnarray}
H^{2}\left( t\right) &=&{\rho \left( t\right) \over 3M_{pl}^{2}}+{%
\Lambda \over 3}-{k \over a^{2}\left( t\right) } \\
\dot{H}\left( t\right) &=&{dH\left( t\right) \over dt}=-{1 \over 2M_{\rm pl}^{2}%
}\left( \rho \left( t\right) +p\left( t\right) \right) +{k \over a^{2}\left(
t\right) }
\end{eqnarray}%
where the Hubble parameter $H(t)$ is defined as%
\begin{equation}
H(t) ={\dot{a}\left( t\right) \over a\left( t\right) }
\end{equation}%
and%
\begin{equation}
M_{\rm pl}=\mbox{\mbox{reduced Planck mass}}=\left( 8\pi G\right) ^{-1/2}.
\end{equation}%
From the energy-momentum conservation condition $\nabla _{\mu }T^{\mu 0}=0$
we find%
\begin{equation}
\dot{\rho}+3H\left( \rho +p\right) =0.
\end{equation}%
This equation and the Friedmann equations are not independent from each
other. The equation of state, the pressure-to-energy density ratio, is
defined as%
\begin{equation}
w={p \over \r }.
\end{equation}%
Typical values of $w$ are%
\begin{eqnarray}
\mbox{Radiation}\mbox{:\ \ \ } &&w=\frac{1}{3} \\
\mbox{Matter}\mbox{:\ \ \ } &&w=0 \\
\mbox{Cosmological constant}\mbox{:\ \ \ } &&w=-1.
\end{eqnarray}

The condition for the accelerating expansion of Universe can be given in several
ways:%
\begin{eqnarray}
\mbox{the cosmic scale factor is accelerating} &\mbox{:}&\mbox{\ \ \ \ \ }
\ddot{a}>0   \nonumber \\
\mbox{the comoving Hubble length is decreasing} &\mbox{:}&\mbox{\ \ \ \ }
{d\over dt}\left({H^{-1}(t) \over a(t) }\right)<0  \label{inf}
\\
\mbox{negative Pressure\ } &\mbox{:}&\mbox{\ \ \ }\rho \left(t\right) +3p\left( t\right) <0. \nonumber
\end{eqnarray}%
The last condition can be also expressed by using the equation of state:%
\begin{equation}
w<-\frac{1}{3}.
\end{equation}

\subsection{Single-Field Inflation Model}

The inflation models assume an era of inflation during the Big Bang, in
which the energy density of the Universe was dominated by the potential
energy of the scalar fields (inflatons). The inflation is defined as any
epoch during which the conditions Eq.\ref{inf} are satisfied and 
used for the events in early Universe. The negative pressure condition for
the inflation model motivates one to introduce a scalar field (inflaton) with
its Lagrangian density given by%
\begin{equation}
\mathcal{L}_{\phi }=\frac{1}{2}g^{\mu \nu }\left( \partial _{\mu }\phi
\right) \left( \partial _{\nu }\phi \right) -V\left( \phi \right) .
\label{Intro028}
\end{equation}%
The term $V\left( \phi \right) $ is the potential of the scalar field and
the choice of its form characterizes the types of inflationary models. Using
the above Lagrangian density we have%
\begin{equation}
T^{\mu \nu }=\left( \partial ^{\mu }\phi \right) \left( \partial ^{\nu }\phi
\right) -g^{\mu \nu }\left[ \frac{1}{2}\left( \partial ^{\lambda }\phi
\right) \left( \partial _{\lambda }\phi \right) +V\left( \phi \right) \right]
\label{Intro029}
\end{equation}%
or%
\begin{equation}
\rho =\frac{1}{2}\dot{\phi}^{2}+\frac{1}{2}\left( \nabla_{i}\phi
\right) ^{2}+V\left( \phi \right)   \label{Intro030}
\end{equation}%
and%
\begin{equation}
p=\frac{1}{2}\dot{\phi}^{2}-\frac{1}{6}\left( \nabla_{i}\phi \right)
^{2}-V\left( \phi \right) .  \label{Intro031}
\end{equation}%
The equation of motion for the scalar field is given by%
\begin{equation}
\ddot{\phi}+3H\dot{\phi}-{1 \over a^{2}\left( t\right) }\vec{\nabla}%
^{2}\phi =-{\partial V \over \partial \phi }.  \label{Intro032}
\end{equation}

\subsection{Quintessence}

In quintessence $Q$ is a scalar field that is introduced to explain the
accelerating Universe. Quintessence is a dynamical field and generally has a
time-varying, spatially inhomogeneous negative pressure [\ref{Ca}]. The
Lagrangian for the quintessence is given by%
\begin{equation}
\mathcal{L}=\frac{1}{2}\partial _{\mu }Q\partial ^{\mu }Q-V\left( Q\right) .
\end{equation}%
The energy density and pressure of quintessence are, then, respectively, 
\begin{equation}
\rho =\frac{1}{2}\dot{Q}^{2}+V\left( Q\right) ,\ \ \ \ p=\frac{1}{2}\dot{Q}%
^{2}-V\left( Q\right) .
\end{equation}%
It is easy to see that the ratio of the pressure to the energy density, $w$,
is $-1<w\leq 0$ for quintessence. A particular class of quintessence
models\thinspace have a tracker behavior. In these models the density of
quintessence closely tracks the radiation density until the equality between
matter and radiation is established. Once the matter-radiation equality is
established the quintessence start to behave as dark energy. k-essence and
phantom energy are the special cases of quintessence.

\subsection{k-essence}

The idea of k-essence (or k-inflation) was originally introduced as a
possible model for inflation [\ref{ADM}]. In this model an inflationary
evolution of the early universe is driven by non-quadratic kinetic energy
terms starting from rather arbitrary initial conditions instead of the
potential energy term $V\left( \varphi \right) $. When the idea of k-essence
is applied to the coincidence problem\footnote{%
This is the problem related to the precision required \thinspace for the
initial value of the total energy density of the Universe so that the
spatial curvature of the Universe is almost zero: the flatness problem. },
it was shown that [\ref{AM}, \ref{AMS}] k-essence can act as a dynamical
attractor at the onset of matter domination period and introduce the cosmic
acceleration at present time. A purely kinetic k-essence model was
investigated in [\ref{Sc1}]. In this investigation the author showed that
k-essence in this model can evolve like the sum of a dark matter component
and a dark energy component if the Lagrangian for the k-essence has a local
extremum.

The Lagrangian for a typical k-essence model is given as the pressure of the
k-essence:%
\begin{equation}
\mathcal{L}=P=V\left( \phi \right) F\left( X\right) ,
\end{equation}%
where $\phi $ is the scalar field representing k-essence, and $X$ is defined
as%
\begin{equation}
X=\frac{1}{2}\partial _{\mu }\phi \partial ^{\mu }\phi .
\end{equation}%
The pressure in this model is given by the Lagrangian itself, while the
energy density is given by%
\begin{equation}
\rho =V\left( \phi \right) \left[ 2XF_{X}-F\right]
\end{equation}%
where%
\begin{equation}
F_{x}={dF\over dX}.
\end{equation}%
Using these pressure and energy density, we have the ratio of the pressure
to the energy density, $w$, as%
\begin{equation}
w={F \over 2XF_{X}-F}.
\end{equation}%
The behavior of this ratio is determined by the form of $V\left( \phi
\right) $ and $F\left( X\right) $.

\subsection{Phantom Dark Energy}

The phantom dark matter is a special case of quintessence characterized by
the ratio of the pressure to the energy density, $w<-1$. Though the dark
energy with $w<-1$ will violate all energy conditions [\ref{Wa}], the dark
energy models with $w<-1$ that are consistent with observed data can be
constructed [\ref{Ca1}]. A phenomenological model of phantom dark energy can
be constructed to give  new insight to the coincident problem through the
interaction of the phantom dark energy with dark energy and dark matter [\ref%
{CW}].

\medskip
\section{The Origin of the Diffeomorphism Field}

In this work we posit that there exist a scalar field, the diffeomorphism scalar field, that has its origins in one of the most primitive symmetries used in physics; general coordinate covariance.  When coupled to a Robertson-Walker metric, this scalar field qualitatively exhibits the behavior of an inflaton, phantom dark energy and the cosmological constant [\ref{RY}].   
Here we review the construction of the Lagrangian for the diffeomorphism field. We will marry the general coordinate transformations with gauge transformations to show the analogy with the vector potential in Yang-Mills theories.  

Recall that under a general coordinate transformation, tensors transform as
\begin{eqnarray}
A'_a(x')	&=& \left({\partial x^b \over \partial x'^a}\right) A_b(x) \\
A'^a(x')	&=& \left({\partial x'^a \over \partial x'^b}\right) A^b(x).
\end{eqnarray}
Their infinitesimal counterpart, i.e. $x'^a=x^a+\xi^a$, where $\xi$ is vector field,  determine the Lie derivative;
\begin{eqnarray}
	{\cal L}_\xi A_a(x) &=& -\xi(x)^b \partial_b A_a(x) - A_b(x)\partial_a \xi^b(x) \\
	{\cal L}_\xi A^a(x) &=& -\xi(x)^b \partial_b A^a(x) + A^b(x)\partial_b \xi^a(x).
\end{eqnarray}
It is important to note that for tensors the Lie derivative is independent of the connection, $\nabla_a$. Thus the field theory associated with diffeomorphisms will be independent of Riemannian curvature.  

The analogue of the above for finite and infinitesimal transformations for gauge transformations on a (non-Abelian) vector potential are 
\begin{equation}
	V'_a(x) = U(x) V_a(x) U(x)^{-1} + i U(x)\partial_a U(x)^{-1}
\end{equation}
and
\begin{equation}
\delta_\Lambda V_a(x) = i \left[\Lambda(x), V_a(x)\right] - \partial_a \Lambda(x),
\end{equation}
where $U(x)=\exp{i \Lambda(x)}$ is the finite gauge transformation generated by $\Lambda(x)$.

\subsection{The Adjoint Representation}

The vector fields ($\xi^a$) and gauge parameters ($\L$) act on themselves and form the adjoint representation.  The gauge and coordinate transformations in one dimension are particularly interesting because the {\em dual} representation of the adjoint representation, called the coadjoint representation, reveals the gauge and gravitational anomalies that appear in two dimensional field theories and string theories.  Interestingly enough, the coadjoint representation appears to come from a field theory that need not be constrained to two dimensions.  In the case of the gauge transformations, that field theory is Yang-Mills.  
  
To see how these field theories comes about, lets begin by marrying the one dimensional infinitesimal coordinate transformations (Lie derivatives) and infinitesimal gauge transformations together.  This marriage is the semi-direct product of the Virasoro algebra [\ref{Vi1}], which corresponds to Lie derivatives of one dimensional vector fields, and an SU(N) Kac-Moody algebra, which corresponds to SU(N) gauge transformations on a circle or a line.  This algebra admits central extensions (or phases at the group level) so that the basic elements of this algebra contains the Lie derivative generated by the one dimensional vector field $\xi^a(\theta)$, a gauge parameter $\Lambda(\theta)$, and a constant, say $a$ so we can have a three-tuple $\left({\xi}; \L; a \right)$.
Then the commutation relations, $[[*,*]]$ for the adjoint representation is 
\begin{equation}
 [[({\xi}; \,\L; \,a), ({\eta}; \X; \,b) ]] =({\xi \circ
   \eta}; \,- \xi^b \partial_b \X + \eta^b \partial_b \L + [\L,\X];\, \left\{\L,\X,\xi,\eta\right\}),
\end{equation}
where the central extensions $\left\{\L,\X,\xi,\eta\right\}$ is given by
\begin{eqnarray}
\left\{\L,\X,\xi,\eta\right\}&=&\frac{k}{2\pi} \int (\L
\partial_a \X - \X \partial_a \L )d\,\theta^a + 
\frac{c}{2\pi} \int (\xi^a \nabla_a \nabla_b \nabla_c
 \eta^c)\,d\theta^b \,\,\,\,\\
 &+& \frac{h}{2\pi} \int (\xi^a \nabla_a \eta_b)\, d\theta^b - (\xi
 \leftrightarrow \eta). \nonumber
 \end{eqnarray}
The parameter $h$,$k$, and $c$ are arbitrary constants that define the central extension.  The new vector field $(\xi \circ \eta)^a$ is defined through the Lie derivative,
\begin{equation}
(\xi \circ \eta)^a \equiv {\cal L}_{\xi} \eta^a = -\xi^b \partial_b \eta^a + \eta^b \partial_b
\xi^a.
\end{equation}

\subsection{The Coadjoint Representation}

Associated with the adjoint representation, we can define its dual by writing an invariant product [\ref{Ki1},\ref{Wi2}] .  Let $\left(\eta; \X; \,b\right)$ be an element of the adjoint and ${\cal B}=\left( D, A, \mu \right)$ be an element of the algebra's dual.  Then the one dimensional line integral gives an invariant pairing:
\begin{equation}
\left\langle \left( D, A, \mu \right) \mid \left(\eta; \X; \,b\right) \right\rangle = \int\left(D_{a b}(\theta) \xi^a(\theta) +A_b(\theta) \L(\theta)\right)\,d\theta^b + b \,\mu.
\end{equation}

Demanding that the pairing to be invariant under gauge and coordinate transformations is equivalent to acting on the pairing with an adjoint element and getting zero.  Consider the action of the adjoint element ${\cal F}=\left( \xi \left( \theta \right) ,\Lambda
\left( \theta \right) ,a\right) $ on the pairing.  Then the invariance of the pairing \begin{equation}
\delta_{\cal F} \left\langle \left( D, A, \mu \right) \mid \left(\eta; \X; \,b\right) \right\rangle = 0
\end{equation} and Leibnitz rule implies that 
\begin{equation}
\delta _{\mathcal{F}}B=\left( \delta_{\cal F}
D\left( \theta \right) ,\delta_{\cal F} A\left( \theta \right) ,0\right)
\end{equation}
where
\begin{equation}
\delta_{\cal F} {\rm D}\left( \theta \right) =\;\stackrel{\rm shift\; in\; the \;diff\;coadjoint\; element}{%
{}{}{\overbrace{\stackunder{coordinate\ trans}{\underbrace{2\xi ^{^{\prime
}}{\rm D}+{\rm D}^{^{\prime }}\xi + \frac {c \mu}{2\pi}\xi^{\prime \prime
\prime } + \frac {h \mu}{2 \pi} \xi^{\prime}}}-%
\stackunder{gauge\ trans}{\underbrace{Tr\left( {\rm A}\Lambda^{\prime }\right) 
}}}}} \end{equation}
and
\begin{equation}
\delta_{\cal F} {\rm A(\theta )}=\;\stackrel{\rm shift\; in\; the\; gauge\; coadjoint\; element}{\,\overbrace{%
\stackunder{ coord\ trans}{\underbrace{{\rm A}^{\prime }\xi +\xi^{\prime }{\rm A}}}-\stackunder{ gauge\ trans}{\,\underbrace{[\Lambda
\,{\rm A}-{\rm A\,}\Lambda ]+k\,\mu \,\Lambda^{\prime }}}}}.
\end{equation}
We have suppressed the tensor indices for the fields $D_{a b}$ and $A_b$ and 
$^{\prime }$ denotes $\partial _\theta$.\\

\subsection{The Geometric Action}

The fields $A$ in the coadjoint
representation correspond to the spatial components of a gauge field
while $D$ is the
space-space component of a rank two tensor called the diffeomorphism
field. The ``$\m$'' is the central extension.  Now there are two distinct Lagrangians that one can construct using the structure of the coadjoint representation.  The first is called the {\em Geometric Lagrangian}.  

In the geometric Lagrangian or geometric action [\ref{Ki3}, {\ref{PS1}}]
the coadjoint elements serve as background fields while the coordinate and group transformations become the dynamical fields.  There one considers a two parameter family of coordinate and gauge transformations that can act on a fixed element of the coadjoint representation, say ${\cal B}=\left(
D\left( \theta \right) ,A\left( \theta \right) ,\tilde{\mu}\right)$.  The two parameters, say $\lambda$ and $\tau$, correspond to a natural coordinatization of the {\em orbit} of ${\cal B}$.  The orbit of ${\cal B}$ is defined to simply be all gauge fields and diffeomorphism fields that are related to ${\cal B}$ via gauge and coordinate transformation.  

On the orbit of any coadjoint element, there exist a natural symplectic structure [\ref{Ar2}] corresponding to a rank two antisymmetric tensor [\ref{Ki3},\ref{Ki4}].  When integrated on a two-manifold this will give the {\em geometric action}.
To find the geometric action for our case, namely the semi-direct product of the Virasoro and
Kac-Moody groups, we define two adjoint vectors corresponding to changes in directions corresponding to coordinates $\l$ and $\t$ on the orbit.  These adjoint vectors are  $u_{\lambda}$ and $u_{\tau}$.  Now let $g$ represent the group action of both an arbitrary gauge transformation, $U(\theta, \l,\t)$, and coordinate transformation, $s(\theta,\l,\t)$.  We denote by $Ad(g)$ the action of these gauge and coordinate transformations of the adjoint representation and  $Ad^{\ast }(g)$ the group action on coadjoint elements.  Then using two adjoint elements corresponding to changes in the $\l$ and $\t$ directions and a coadjoint representation $w_{g}$ defined by
\begin{eqnarray}
u_{\lambda } &=&\left( \xi _{\lambda }\left( \theta \right) , \Lambda
_{\lambda }\left( \theta \right) , a_{\lambda }\right)  \nonumber \\
u_{\tau } &=&\left( \xi _{\tau }\left( \theta \right) , \Lambda _{\tau
}\left( \theta \right) , a_{\tau }\right)  \\
w_{g} &=&Ad^{\ast }\left( g\right) w = Ad^{\ast }\left( g\right) \left(
D\left( \theta \right) , A\left( \theta \right) , \tilde{\mu}\right).  \nonumber
\end{eqnarray}%
We then pair these elements via  $\left\langle w_{g}\ |\ \left[ u_{\lambda },u_{\tau }\right]
\right\rangle $ to define a natural symplectic two form in the coordinates of $\l$ and $\t$.  This is given by%
\begin{equation}
\left\langle w_{g}\ |\ \left[u_{\lambda },u_{\tau }\right] \right\rangle
=\left\langle Ad^{\ast }\left( g\right) w\ |\ \left[u_{\lambda },u_{\tau }%
\right]\right\rangle =\left\langle w\ |\ \left[ Ad\left( g^{-1}\right)
u_{\lambda },Ad\left( g^{-1}\right) u_{\tau }\right] \right\rangle
\label{VKM024}
\end{equation}%
Concretely we have%
\begin{equation}
\xi _{\lambda }\left( \theta \right) =\left( \partial _{\lambda }s\left(
\theta \right) \right) \circ s^{-1}\ \ \ \ \ \ \mbox{and}\ \ \ \ \ \ \Lambda
_{\lambda }\left( \theta \right) =\left( \partial _{\lambda }U\right) \circ
U^{-1},  \label{VKM025}
\end{equation}%
and we find%
\begin{eqnarray}
Ad\left( g^{-1}\right) \xi _{\lambda }\left( \theta \right) &=&s^{-1}\circ
\left( \partial _{\lambda }s\left( \theta \right) \right) =\partial
_{\lambda }\theta =\frac{\partial _{\lambda }s}{\partial _{\theta }s}
\label{VKM026} \\
Ad\left( g^{-1}\right) \Lambda _{\lambda }\left( \theta \right)
&=&U^{-1}\circ \left( \partial _{\lambda }U\right) =U^{-1}\partial _{\lambda
}U. \label{VKM027}
\end{eqnarray}%
Then in Eq.(\ref{VKM024}) we can write
\begin{eqnarray}
&&\left[ Ad\left( g^{-1}\right) u_{\lambda },Ad\left( g^{-1}\right) u_{\tau }%
\right] = \nonumber \\
&&\left[ \left( \frac{\partial _{\lambda }s}{\partial _{\theta }s}%
,U^{-1}\partial _{\lambda }U,a_{\lambda }\right) ,\left( \frac{\partial
_{\tau }s}{\partial _{\theta }s},U^{-1}\partial _{\tau }U,a_{\tau }\right)%
\right]  =\\
&&\left( \left[ \frac{\partial _{\lambda }s}{\partial _{\theta }s},\frac{%
\partial _{\tau }s}{\partial _{\theta }s}\right],\left[ \frac{\partial
_{\lambda }s}{\partial _{\theta }s},U^{-1}\partial _{\tau }U\right] +\left[
U^{-1}\partial _{\lambda }U,\frac{\partial _{\tau }s}{\partial _{\theta }s}%
\right] +\left[ U^{-1}\partial _{\lambda }U,U^{-1}\partial _{\tau }U\right]
, c_{V}+c_{K}\right) \nonumber
\end{eqnarray}%
where $c_{V}$ and $c_{K}$ are the central terms due to the commutation
relations for the Virasoro and Kac-Moody algebras, respectively. Choosing $c=-h$ [\ref{KR}] ,
the central term for the Virasoro commutator is given by%
\begin{equation}
c_{V}=\frac{c}{24\pi }\int_{0}^{2\pi }d\sigma u^{\prime }v^{\prime \prime }
\label{VKM029}
\end{equation}%
where $u=\frac{\partial _{\lambda }s}{\partial _{\theta }s}$ and $v=\frac{%
\partial _{\tau }s}{\partial _{\theta}s}$, whereas the Kac-Moody commutator
gives%
\begin{equation}
c_{k}=\frac{k}{2\pi }\int_{0}^{2\pi }d\sigma Tr\left( u v'\right) ,  
\end{equation}%
where $u=U^{-1}\partial _{\lambda }U$ and $v=U^{-1}\partial _{\tau }U$.
Hence, after some calculations, the geometric action 
\begin{equation}
S_{V\otimes KM} 
=\int d\lambda d\tau \left\langle w\ |\ \left[ Ad\left( g^{-1}\right)
u_{\lambda },Ad\left( g^{-1}\right) u_{\tau }\right] \,\right\rangle
\end{equation}
becomes [\ref{AS},\ref{RR},\ref{LR1},\ref{LR2}]
\begin{eqnarray}
&S& =\frac 1{2\pi }\stackunder{\mbox{\small{Metric \,Coupling}}}
{\underbrace{%
\int d^{\;2}\theta \;{\rm D}\left( \theta \right) \left( \frac{%
\partial _\tau s}{\partial _\theta s}\right)  }}  \nonumber \\
&+&\frac 1{2\pi }{%
\int d^{\;3}\theta \;\mbox{Tr} {\rm A}\left( \theta \right)
\left( \frac{\partial _\lambda s}{\partial _\theta s}\partial _\theta \left(
U^{-1}\partial _\tau U\right) \right)} \label{VKM031} \\
&-&\frac 1{2\pi }\int d^{\;3}\theta \;\mbox{Tr} {\rm A}(\theta){\left(\frac{\partial _\tau s}{\partial _\theta s}%
\partial _\theta \left( U^{-1}\partial _\lambda U\right) +\left[
U^{-1}\partial _\lambda U,U^{-1}\partial _\tau U\right] \right)  } 
\nonumber \\
&-&{\beta c \over 48\pi }
{\int
d^{\;2}\theta \;\left( {\partial _\theta ^2s \over \left( \partial _\theta
s\right) ^2}\partial _\tau \partial _\theta s-{\left( \partial _\theta
^2s\right) ^2 \over\left( \partial _\theta s\right) ^3}\partial _\tau s\right)
 \label{3.7} } \nonumber \\
&-&{\beta k \over 4\pi }
{\int d^{\;3}\theta \;{\rm Tr}\left( U^{-1}\partial _\theta UU^{-1}\partial _\tau
U\right) } \nonumber \\
&+&{{\beta k \over 4\pi }\int d^{\;3}\theta \;\mbox{Tr}\left(
\left[ U^{-1}\partial _\theta U,U^{-1}\partial _\lambda U\right]
U^{-1}\partial _\tau U\right) \nonumber },
\end{eqnarray}
where the measures $d^{\;2}\theta=d\theta d\tau$ and $d^{\;3}\theta=d\theta d\tau d\lambda$.

The fourth summand is recognized as the Liouville-Polyakov action [\ref{P1}] while the fifth and sixth terms are the Wess-Zumino-Novikov-Witten actions [\ref{Wi1}].  In the summand called the ``Metric Coupling", one sees that the diffeomorphism field interacts with the metric through 
\[ 
{\rm D}\left( \theta \right) \left( \frac{
\partial _\tau s}{\partial _\theta s}\right) \rightarrow \sqrt{h} D_{a b}h^{a b},\] where $h_{ab}$ is the two dimensional induced metric.   {\em This interaction is the first ingredient we will need to discuss cosmology.}  When the trace of $D_{ab}$ becomes constant, this term in the Lagrangian will appear as a cosmological constant.
  
\subsection{Transverse Actions}
Every point on the coadjoint orbit of ${\cal B}=\left( D, A, \mu \right)$ corresponds to a group element, $s(\theta)$ and $U(\theta)$.  There are, however, group elements that do not appear on the orbit because they do not change ${\cal B}=\left( D, A, \mu \right)$.  These group elements are generated by the {\em isotropy algebra} of the coadjoint element, ${\cal B}=\left( D, A, \mu \right)$.  An adjoint element ${\cal F}$ belongs to the isotropy algebra of ${\cal B}=\left( D, A, \mu \right)$ when
\begin{equation}
\delta_{\cal F} {\rm D}\left( \theta \right) =\;{%
{}{}{2\xi ^{^{\prime
}}{\rm D}+{\rm D}^{^{\prime }}\xi + \frac {c \mu}{2\pi}\xi^{\prime \prime
\prime } + \frac {h \mu}{2 \pi} \xi^{\prime}-
{{\rm Tr}\left( {\rm A}\Lambda^{\prime }\right) 
}}}=0\end{equation}
and
\begin{equation}
\delta_{\cal F} {\rm A(\theta )}=\;{\rm A}^{\prime }\xi +\xi^{\prime }{\rm A}-{[\Lambda
\,{\rm A}-{\rm A\,}\Lambda ]+k\,\mu \,\Lambda^{\prime }}=0.
\end{equation}
These adjoint elements are candidates for the conjugate momenta of the fields $D$ and $A$ that appear in ${\cal B}=\left( D, A, \mu \right)$.  To see this recall that in Yang-Mills theories the finite gauge transformations are given by%
\begin{eqnarray}
A_{i}\left( x\right) &\rightarrow &U\left( x\right) A_{i}\left( x\right)
U^{-1}\left( x\right) -\left( \partial _{i}U\left( x\right)
\right) U^{-1}\left( x\right)  \label{YM001} \\
E^{i}\left( x\right) &\rightarrow &U\left( x\right) E^{i}\left( x\right)
U^{-1}\left( x\right) .  \label{YM002}
\end{eqnarray}%
In temporal gauge%
\begin{equation}
A_{0}=0,  \label{YM003}
\end{equation}
the residual gauge transformation are time-independent gauge
transformations. The infinitesimal residual gauge transformations are:%
\begin{eqnarray}
\delta A_{ai}\left( x\right) &=&\left[ \Lambda \left( x\right) ,A\left(
x\right) \right] _{ai}-\partial _{i}\Lambda _{a}\left( x\right)
\label{YM004} \\
\delta E_{a}^{i}\left( x\right) &=&\left[ \Lambda \left( x\right) ,E\left(
x\right) \right] _{a}^{i}.  \label{YM005}
\end{eqnarray}
The \textit{isotropy algebras} are given by setting these equations to zero:%
\begin{eqnarray}
\delta_\Lambda A_{ai}\left( x\right) &=&\left[ \Lambda \left( x\right) ,A\left(
x\right) \right] _{ai}-\partial _{i}\Lambda _{a}\left( x\right) =0
\label{YM006} \\
\delta_\Lambda E_{a}^{i}\left( x\right) &=&\left[ \Lambda \left( x\right) ,E\left(
x\right) \right] _{a}^{i}=0.  \label{YM007}
\end{eqnarray}%

In Eq.(\ref{YM001}) and Eq.(\ref{YM002}), the transformation laws reveal that in one space and one time dimensions, the gauge potential is in the coadjoint representation while its conjugate momentum, the electric field, is in the adjoint representation.  Because $E_i$ is in the adjoint algebra it can gauge transform $A_i$.  The  Gauss law constraints%
\begin{equation}
\nabla _{i}E_{a}^{i}\left( x\right) =\partial _{i}E_{a}^{i}\left( x\right)
+ \left[ A_{i}\left( x\right) ,E^{i}\left( x\right) \right] _{a}=0,
\label{YM008}
\end{equation}%
forbids this from happening.  From Eq.(\ref{YM006}) we see that the {\em conjugate momentum to the coadjoint element is an element of the isotropy algebra}.  With this one now has a way of finding a Lagrangian that will give dynamics to the fields in ${\cal B}$.  We construct a Lagrangian where one of the field equations reduces to a Gauss Law constraint which corresponds to the isotropy equation in two dimensions.  The Lagrangian can exist in any dimension.  For the Kac-Moody sector the Lagrangian is the familiar: 
\begin{equation}
S_{YM}=-\frac{1}{2}\int d^{n}x\;{\rm Tr}\left( F_{\mu \nu }\left( x\right) F^{\mu
\nu }\left( x\right) \right).  \label{YM021}
\end{equation}%

\begin{figure}
\centering
  \includegraphics[scale=.6]{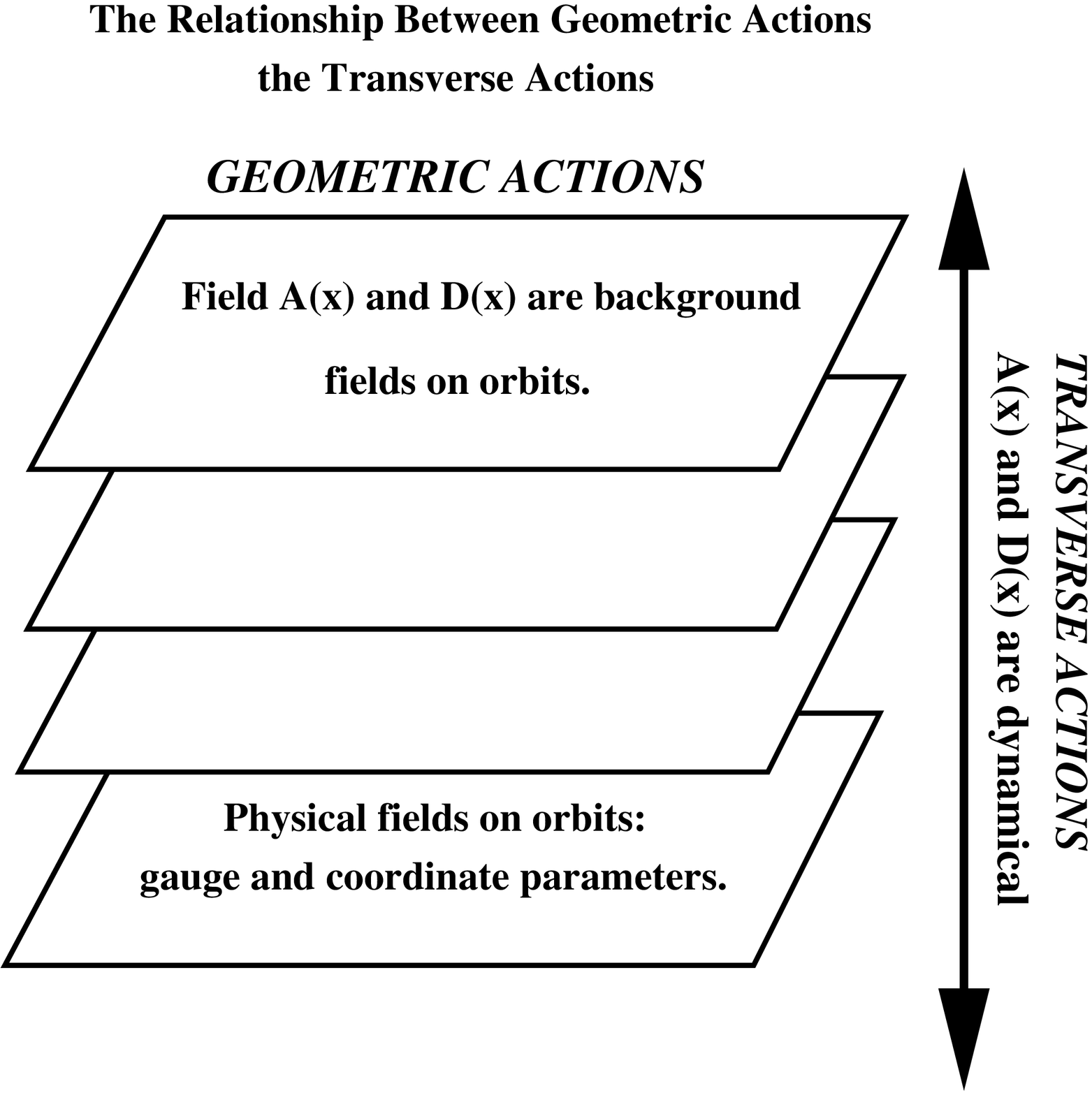}
  \caption{The Foliation of the Dual of the Algebra}
  \label{fig0}
\end{figure}

This approach was applied to the isotropy equation obtained from the
Virasoro algebra,
\begin{equation}
\delta _{\xi }D=2D^{\prime }\xi +D\xi ^{\prime }+q\xi ^{\prime \prime \prime
}+\beta \xi ^{\prime }=0  \label{TAV002}
\end{equation}%
where%
\begin{equation}
q=\frac{c\mu }{2\pi }\ \ \ \ \mbox{and\ \ \ \ }\beta =\frac{h\mu }{2\pi }.
\end{equation}%
Following the Yang-Mills case, we assume that this equation is obtained
after enforcing the temporal gauge and on finds the {\em transverse action},
\begin{eqnarray}
S_{\mbox{\tiny diff}}=&-&\int d^nx \sqrt{g}~\alpha
\left( X^{l m r}~{\rm D}^a{}_r 
X_{m l a} +2 X^{l m r} {\rm D}_{l a}
X^a{}_{r m}\right)\\
  &-&\int d^nx \sqrt{g}\left(q  
X^{a b}{}_b {} \nabla_l \nabla_m{} X^{l m }{}_a+ \frac{\b}2  X^{b g a} 
X_{b g a}-\frac{1}{2}X_{m n r }\left( x\right) X^{m n r }\left(
x\right) \right) \nonumber
\end{eqnarray}
where ${\rm X}^{m n r} = \nabla^r D^{m n}$. The term proportional to $q$ arises from the central extension in the algebra.  Algebraically the central extension exists only in one dimension.  However, in the construction of the transverse action, this contribution is seen as coming from a derivative interaction of the connection with the diffeomorphism field.  We expect this term to be relevant in the early universe when there is only time dependence in the system.  As we will see later, after time evolution, the $q$ dependence of the system becomes negligible.  This could be the signature that the symmetries in the cosmological principle are broken and that spatial dependence becomes relevant.  

A pictorial interpretation of the geometric and transverse action can be seen in Fig.(\ref{fig0}). The geometric actions is the physics of the coordinates and gauge fields (collective coordinates) and is represented by horizontal slices on the foliation of the dual of the algebra.  These horizontal slices are equivalence classes of fields and any physics there would constitute anomalies in the gauge and diffeomorphism symmetries.  One the other hand, the transverse actions move the fields from one equivalence class to another and do not contain spurious gauge and coordinate degrees of freedom because of the Gauss law constraints.  It is for this reason that we say that the transverse actions are transverse to the geometric actions.   
In $n$ dimensions the mass dimensions for the coupling constants 
and the diff field are:%
\begin{equation}
\left[ \alpha \right] =\frac{2-n}{2},\ \ \ \left[ \beta \right] =0,\ \ \ %
\left[ q\right] =-2,\ {\mbox{\rm and}}\ \ \ \left[
D_{\mu \nu }\right] =\frac{n-2}{2}.
\end{equation}

To see that the diffeomorphism action recovers Eq.(\ref{TAV002}) in two dimensions, we 
vary the action with respect to the space-time component ${\rm D}_{i 0}$. Then by setting ${\rm D}_{i 0}=0$, the field equation becomes
\begin{equation}
X^{l m 0} \partial_i {\rm D}^{l m} - \partial_m (X^{m l 0} {\rm D}_{l i}) -\partial_l(X^{m l 0} {\rm D}_{m i}) - 
q{~}\partial_i \partial_l \partial_m X^{l m 0}= 0. \label{diffeq1}
\end{equation}
In $1+1$ this corresponds to the isotropy equation found on the coadjoint
orbit where D corresponds to to space-space component, $D_{11}$.  The field
equations in 1+1 dimensions reduce to 
\begin{equation}
X {\rm D}' + 2 X' {\rm D} + q{~} X''' =0 \,\, \rightarrow \,\, 
\xi  {\rm D}' + 2 \xi' {\rm D} + q{~} \xi''' =0,
\end{equation}
 where the adjoint element
$\xi$ corresponds to the conjugate momentum, $X \equiv X^{1 1 0}$, of  ${\rm
  D}\equiv{\rm  D}_{1 1}$.  This action is exactly analogous to Yang-Mills
theory.
\newpage
\section{\protect\medskip Cosmology and the Diffeomorphism
Field}

We have seen that there are two types of physical action naturally arise
from the coadjoint representations of the Virasoro algebras
(diffeomorphisms) and Kac-Moody algebra (affine Lie algebra). One
action is the geometric action that comes from the coadjoint orbits of
algebras. The other action is from the phase space that is transverse to the
coadjoint orbits: the transverse action. The transverse action of the
Kac-Moody algebra is the Yang-Mills action in two dimensions and it is also
valid in higher dimensions. The transverse action for the Virasoro algebra
can be constructed by following the analogy of the construction for the
Yang-Mills theory. The action associated with the algebra of diffeomorphisms is also
valid in higher dimensions as its origins are Lie Derivatives.  

The Virasoro algebra is the symmetry under coordinate
transformation on the unit circle, i.e., diffeomorphisms. There an important
lesson from string theory is that diffeomorphism should play an active role in
the theory of gravitation. Indeed, we have already seen important
implication of diffeomorphism in the geometric action for the Virasoro
algebra Eq.[\ref{VKM031}]:%
\begin{equation}
\frac{1}{2\pi }\int d\tau d\theta D\left(\theta \right) \left(\frac{
\partial _{\tau }s}{\partial _{\theta }s}\right).  \label{C001}
\end{equation}%
If we rewrite this term in coordinate invariant notion, we have%
\begin{equation}
\int \sqrt{h}D_{ab}h^{ab}dx  \label{C002}
\end{equation}%
where $h^{ab}$ is the inverse metric on the two manifold. Our interpretation
of the diffeomorphism field $D$ is that it is a non-Riemannian contribution to
gravitation: it appears already in two dimensions where the Einstein-Hilbert
action only offers topological information. It is easy to observe that when
the trace of $D_{ab}$ takes a constant value, this term appears as a natural
source for the cosmological constant. Such a term in the action does not
depend on the two dimensional structure.

In this section we will apply the transverse action of the Virasoro algebra
to cosmology. More specifically, we will study the coupling of the trace
part of the diffeomorphism field $D$ to the Robertson-Walker metric for the
flat Universe ($k=1$) and study the cosmological implications of the diff
field. Many of the calculations in this section are done by Mathematica with
MathTensor package [\ref{PC},\ref{WO}]. 
The action used to study the cosmological effect of the diff field $D$ has
two parts:%
\begin{equation}
S=S_{\mbox{\tiny HE}}+S_{\mbox{\tiny diff}}  \label{C003}
\end{equation}%
where $S_{\mbox{\tiny HE}}$ is the Hilbert-Einstein action without cosmological constant 
and $S_{\mbox{\tiny diff}}$ is the transverse action for diff field:%
\begin{equation}
S_{\mbox{\tiny HE}}=-{1\over 16\pi G}\int dx^{4}\sqrt{g\left( x\right) }R\left( x\right)
,\ \ \ \ \ x=\left( \vec{x},t\right)  \label{C004}
\end{equation}%
and%
\begin{eqnarray}
S_{\mbox{\tiny diff}}=&-&\int d^nx \sqrt{g}~\alpha
\left( X^{l m r}~{\rm D}^a{}_r 
X_{m l a} +2 X^{l m r} {\rm D}_{l a}
X^a{}_{r m}\right) \nonumber \\
  &-&\int d^nx \sqrt{g}\left(q  
X^{a b}{}_b {} \nabla_l \nabla_m{} X^{l m }{}_a+ \frac{\b}2  X^{b g a} 
X_{b g a}\right) \label{C005}\\
&-& \frac{1}{2}\int d^nx \,\sqrt{g}X_{m n r } X^{m n r } + \lambda \int d^nx \sqrt{g}\,D_{\ m }^{m }.\nonumber
\end{eqnarray}
We note that $\alpha $ term is from Gauss' law constraint, $q$ and $\beta $
terms are from the central extension of the algebra. $\lambda $ term is
analogous to Eq.\ref{C002}, i.e., the cosmological term. Since we will set
the ordinary cosmological constant $\Lambda $\thinspace to zero, there will
be no ambiguity. Again the mass dimensions in $n$ dimensions are%
\begin{equation}
\left[ \alpha \right] =\frac{2-n}{2},\ \ \ \left[ \beta \right] =0,\ \ \ %
\left[ q\right] =-2,\ \ \ \left[ \lambda \right] =\frac{n+2}{2},\ \ \ \left[
D_{\mu \nu }\right] =\frac{n-2}{2}.
\end{equation}%
The equations of motion and the energy momentum tensor for the diff field $%
D_{\mu \nu }$ are obtained by varying the diff action $S_{\mbox{\tiny diff}}$ with
respect to $D_{\mu \nu }$ and $g_{\mu \nu }$, respectively:%
\begin{equation}
\mbox{equations of motion:\ \ \ }{\delta S_{\mbox{\tiny diff}} \over \delta D_{\mu \nu }}%
=0
\end{equation}%
\medskip and%
\begin{equation}
\mbox{energy-momentum tensor:\ \ \ }\delta S_{\mbox{\tiny diff}}=\frac{1}{2}\int d^{4}x%
\sqrt{g\left( x\right) }\,T^{\mu \nu }\left( x\right) \delta g_{\mu \nu }.
\end{equation}%

The trace of the diff field $D_{\mu \nu }$ is defined through the
decomposition%
\begin{equation}
D_{\mu \nu }\left( x\right) =\frac{1}{n}\psi \left( x\right) g_{\mu \nu
}+W_{\mu \nu }\left( x\right)
\end{equation}%
where $\psi \left( x\right) $ is a scalar field, $W_{\mu \nu }\left(
x\right) $ is a traceless symmetric tensor field, and $n$ is the dimension
of the spacetime. Since we are interested only in the trace part of the
diff field $D_{\mu \nu }$, we set the traceless part to zero: $W_{\mu \nu
}\left( x\right) =0$. The equations of motion for $D_{\mu \nu }\left(
x\right) $, reduce to [\ref{YT}]
\begin{eqnarray}
&&\left( 4-{4\beta \over n}-2\alpha\left( {2\over n^{2}}+{1\over n}\right)
 \psi \left( x\right) \right) \nabla ^{\mu }\nabla _{\mu }\psi \left(
x\right) \\
&&\ \ \ -\alpha \left( {2\over n^{2}}+{1\over n}\right) \nabla ^{\mu }\psi
\left( x\right) \nabla _{\mu }\psi \left( x\right) -{1\over n}q\nabla ^{\mu
}\nabla ^{\nu }\nabla _{\mu }\nabla _{\nu }\psi \left( x\right) +n\lambda =0.
\nonumber
\end{eqnarray}%
where $\nabla _{\mu }$ is the covariant derivative. The energy-momentum
tensor for $D_{\mu \nu }$ is reduced to [\ref{YT}]
\begin{eqnarray}
T^{\mu \nu } &=&2\left( {2\over n}-{2\beta \over n^{2}}-\left( {1 \over%
n^{2}}+{2 \over n^{3}}\right) \alpha \psi \left( x\right) \right) \nabla
^{\mu }\psi \left( x\right) \nabla ^{\nu }\psi \left( x\right)  \nonumber \\
&&+{q\over n^{2}}\left( 
\begin{array}{c}
\nabla ^{\sigma }\nabla _{\sigma }\psi \left( x\right) \nabla ^{\mu }\nabla
^{\nu }\psi \left( x\right) +\nabla _{\sigma }\psi \left( x\right) \nabla
^{\sigma }\nabla ^{\mu }\nabla ^{\nu }\psi \left( x\right) \\ 
\\ 
-\nabla ^{\mu }\psi \left( x\right) \nabla ^{\sigma }\nabla ^{\nu }\nabla
_{\sigma }\psi \left( x\right) -\nabla ^{\nu }\psi \left( x\right) \nabla
^{\sigma }\nabla ^{\mu }\nabla _{\sigma }\psi \left( x\right)%
\end{array}%
\right)  \nonumber \\
&&-\left( {2\over n}-{2\beta \over n^{2}}-\left({1\over n^{2}}+{2 \over %
n^{3}}\right) \alpha \psi \left( x\right) \right) \nabla ^{\sigma }\psi
\left( x\right) \nabla _{\sigma }\psi \left( x\right) g^{\mu \nu }  \nonumber \\
&&-{q\over 2n^{2}}\nabla ^{\sigma }\nabla ^{\tau }\psi \left( x\right)
\nabla _{\sigma }\nabla _{\tau }\psi \left( x\right) g^{\mu \nu }  \nonumber \\
&&+\lambda \psi \left( x\right) g^{\mu \nu }.
\end{eqnarray}%

Applying the cosmological principle to the diffeomorphism field, we write, in four
dimensions,%
\begin{equation}
D_{\mu \nu }\left( t\right) =\frac{1}{4}f(t) g_{\mu \nu }
\end{equation}%
where $f(t) =\psi \left( \vec{x},t\right) $ is a scalar function
that only depends on time. Using this form of diff field, we find the equation of motion for the
diff field as%
\begin{equation}
\begin{array}{c}
\left( 1-\frac{\beta }{4}-\frac{3}{16}\alpha f\left( t\right) \right)
f^{\prime \prime }\left( t\right) +3H\left[ \left( 1-\frac{\beta }{4}-\frac{3%
}{16}\alpha f\left( t\right) \right) f^{\prime }\left( t\right) +\frac{1}{16}%
qf^{\left( 3\right) }\left( t\right) \right] \\ 
\\ 
+\lambda -\frac{3}{32}\alpha f^{\prime }(t) ^{2} +\frac{1}{16}qf^{\left( 4\right) }(t) +\frac{3}{8} q H^{2}f^{^{\prime \prime }}f\\ 
\\ 
 -\frac{9}{16}qH^{3}f^{\prime
}(t) +\frac{3}{16}qH^{\prime }f^{\prime \prime }(t) -%
\frac{3}{8}qf^{\prime }(t) H^{\prime }H=0%
\end{array}
\label{EQMs01}
\end{equation}%
Here we used ` $^{\prime }$ ' as a time derivative and $H=H\left( t\right) $. The energy density and pressure of the diff field are also found,
respectively, as%
\begin{eqnarray}
\rho _{\mbox{\tiny diff}}\left( t\right) &=&\frac{1}{2}\left( 1-\frac{\beta }{4}-\frac{3}{%
16}\alpha f\left( t\right) \right) f^{\prime }\left( t\right) ^{2}+\lambda
f\left( t\right) \\
&&+\frac{7}{32}qH^{2}f^{\prime }\left( t\right) ^{2}+\frac{1}{8}qHf^{\prime
}\left( t\right) f^{\prime \prime }\left( t\right) -\frac{1}{32}qf^{^{\prime
\prime }}\left( t\right) ^{2}  \nonumber
\end{eqnarray}%
and%
\begin{eqnarray}
p_{\mbox{\tiny diff}}\left( t\right) &=&\frac{1}{2}\left( 1-\frac{\beta }{4}-\frac{3}{16}%
\alpha f\left( t\right) \right) f^{\prime }\left( t\right) ^{2}-\lambda
f\left( t\right) \\
&-&\frac{9}{32}qH^{2}f^{\prime }\left( t\right) ^{2}+\frac{3}{16}qHf^{\prime
}\left( t\right) f^{\prime \prime }\left( t\right) -\frac{1}{32}qf^{^{\prime
\prime }}\left( t\right) ^{2}+\frac{1}{16}qf^{\prime }\left( t\right)
f^{\left( 3\right) }\left( t\right) .  \nonumber
\end{eqnarray}%
Note that the equation of motion for the diff field depends on both $H$ and $%
H^{\prime }$, where as the energy density and the pressure only depend on $H$. The equation of state, the pressure-to-energy density ratio, $w\left(
t\right) $, for the diff field is given by%
\begin{equation}
w\left( t\right) ={p_{\mbox{\tiny diff}}\left( t\right) \over \rho _{\mbox{\tiny diff}}\left(
t\right) }
\end{equation}

To study the properties of the diff field in the early universe, we assume
that the contributions from the other sources are negligible. In other
words, we set the energy densities and pressures for other matters
(including radiation) to zero. Under this assumption the Friedmann
equations can be written in terms of the scalar field $f\left(t\right) $
and its derivatives in time:%
\begin{eqnarray}
H^{2}(t) &=&{1\over 3M_{\rm pl}^{2}}\left[ 
\begin{array}{c}
\frac{1}{2}\left( 1-\frac{\beta }{4}-\frac{3}{16}\alpha f\left( t\right)
\right) f^{\prime }\left( t\right) ^{2}+\lambda f\left( t\right) \\ 
\\ 
+\frac{7}{32}qH^{2}f^{\prime \prime }\left( t\right) +\frac{1}{8}qHf^{\prime
}\left( t\right) f^{\prime \prime }\left( t\right) -\frac{1}{32}qf^{^{\prime
\prime }}\left( t\right) ^{2}%
\end{array}%
\right]  \label{FrieS01} \\
&&  \nonumber \\
H^{\prime }\left( t\right) &=&-{1\over 2M_{\mbox pl}^{2}}\left[ 
\begin{array}{c}
\left( 1-\frac{\beta }{4}-\frac{3}{16}\alpha f\left( t\right) \right)
f^{\prime }\left( t\right) ^{2} \\ 
\\ 
-{1 \over 16}qf^{^{\prime \prime }}\left( t\right) ^{2}+\frac{1}{16}%
qf^{\prime }\left( t\right) f^{\left( 3\right) }\left( t\right) \\ 
\\ 
-\frac{1}{16}qH^{2}f^{\prime }\left( t\right) ^{2}+\frac{5}{16}qHf^{\prime
}\left( t\right) f^{\prime \prime }\left( t\right)%
\end{array}%
\right]  \label{FrieS02} \\
&&  \nonumber
\end{eqnarray}%
The expression of Hubble parameter $H\left( t\right) $ can be obtained from
Eq.\ref{FrieS01}. Since it is in a quadratic form in $H$, we have two
possibilities:%
\begin{equation}
H_{\pm}(t) ={1 \over 96M_{\mbox pl}^{2}+9qf^{\prime }\left( t\right) ^{2}}%
\left[ 3qf^{\prime }\left( t\right) f^{\prime \prime }\left( t\right) \pm 
\sqrt{Q}\right]  \label{Hubs01}
\end{equation}%
where%
\begin{eqnarray}
Q &=&9q^{2}f^{\prime }\left( t\right) ^{2}f^{\prime \prime }\left( t\right)
^{2}  \label{Hubs02} \\
&&-\left( 96M_{pl}^{2}+9qf^{\prime }\left( t\right) ^{2}\right) \left( 
\begin{array}{c}
4\left( \beta -4\right) f^{\prime }\left( t\right) ^{2}+f\left( t\right)
\left( 32\lambda +3\alpha f^{\prime }\left( t\right) ^{2}\right) \\ 
+qf^{\prime \prime }\left( t\right) ^{2}-2qf^{\prime }\left( t\right)
f^{\left( 3\right) }\left( t\right)%
\end{array}%
\right) .  \nonumber
\end{eqnarray}

The Friedmann equations, Eq.\ref{FrieS01} and Eq.\ref{FrieS02}, also allow us
to simplify the equation of motion for $D_{\mu \nu }$, Eq.\ref{EQMs01}:
these two equations can be used to eliminate $H$ and $H^{\prime }$ in the
equation of motion for $D_{\mu \nu }$. 
\medskip

\section{Numerical Results}

\medskip

We will numerically solve the equation of motion for the trace
part of $D_{\mu \nu }$ in the Robertson-Walker metric. We use the numerical
solution of $f\left( t\right) $ to plot the Hubble parameter $H\left(
t\right) $, energy-density $\rho \left( t\right) $, pressure $p\left(
t\right) $, the equation of state $w\left( t\right) $ and $\rho \left(
t\right) +3p\left( t\right) $. Since the equation of motion\thinspace that
we are studying is very complicated, we will limit our analysis to the basic
behavior of each quantities under given set of parameters and initial
conditions. Our main goal in this numerical computation is not to obtain the
results that agree with the current observation, but to investigate the
effects of the trace part of diff field $D_{\mu \nu }$ on the quantities
mentioned above to study the nature of the scalar field $f\left( t\right) $.
We use the explicit Runge-Kutta method for this numerical computation.

\subsection{Parameters}

Eq.\ref{Hubs01} suggests that the Hubble parameter $H\left( t\right) $ is
expected to increase its magnitude very rapidly as%
\begin{equation}
f^{\prime }\left( t\right) ^{2}\rightarrow -\frac{32}{3q}M_{\mbox pl}^{2}.
\end{equation}%
The above condition also implies that the value of $q$ should be negative.
In this limit it is not difficult see that the positive solution (the
solution with $+\sqrt{Q}$) has a simpler behavior near the ``peak". Hence, in
the following, we will study the positive solution case. Noting that the
function of parameter $\beta $ is basically to shift the value of kinetic
term, we set this parameter to zero. Noting that the mass dimension scalar
field $f\left( t\right) $ is one, so it's natural to use the planck mass $%
M_{pl}$ for this scalar field, i.e., we set $M_{pl}$ to one. Hence
undetermined parameters are $\alpha $, $q$, and $\lambda $.

\subsection{Trial Initial Conditions}

\begin{figure}
		\includegraphics[scale=.3,angle=90]{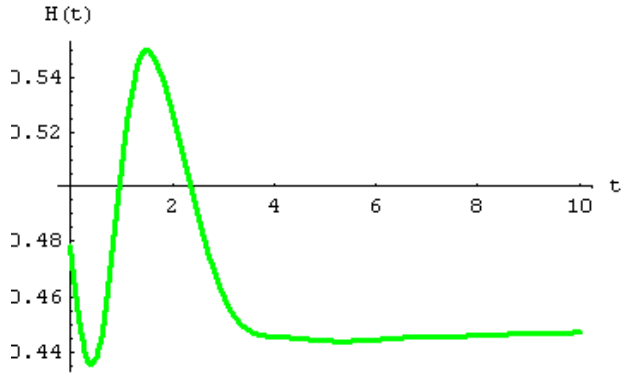}
	\caption{The Trial Hubble Parameter Plot}
	\label{fig:hubble}
\end{figure}

The equation of motion for the trace part of $D_{\mu \nu }$ in the
Robertson-Walker metric is a fourth order nonlinear differential equation
 with three parameters (after setting $\beta =0$ and $M_{pl}=1$). To solve these differential equations numerically, we need to specify trial values for the
three parameters and four initial conditions. In this numerical
computation the initial conditions for $f\left( t\right) $, $f^{\prime
}\left( t\right) $, $f^{\prime \prime }\left( t\right) $, and $f^{\left(
3\right) }\left( t\right) $ are used. A priori, we do not
 know how to choose these parameters and initial conditions. To overcome this, we use the physical condition on
the Hubble parameter as a guide for choosing the trial values. We search
for a set of the parameters and the initial conditions that give a positive
and finite value for the Hubble parameter. 
With this as a criterion, a choice for the trial of parameters and the initial
conditions become
\begin{eqnarray}
&&\alpha =-1,\ \ \ q=-6,\ \ \ \lambda =0.05,  \nonumber \\
f\left( 0\right) &=&-10,\ f^{\prime }\left( 0\right) =0,\ f^{\prime \prime
}(0)=-1,\ f^{\left( 3\right) }=-1.  \label{ref}
\end{eqnarray}%
This set of trial values give the Hubble parameter plot shown in Fig.\ref{fig:hubble}. 
Note that we used$\ k=0$ (flat Universe), $\Lambda =0$ (no cosmological
constant), $\beta =0$, and $M_{\mbox pl}=1$.

\subsection{Computation}

As we have noted in Chapter 1, the nature of scalar fields are characterized
by the equation of state $w\left( t\right) $, and this value is defined as
the ratio of the pressure $p\left( t\right) $ to the energy density $\rho
\left( t\right) $. The acceleration of Universe can be studied through the
quantity $\rho \left( t\right) +3p\left( t\right) $. We, thus, compute $%
f\left( t\right) $, $H\left( t\right) $, $\rho \left( t\right) $, $p\left(
t\right) $, $w\left( t\right) $, and $\rho \left( t\right) +3p\left(
t\right) $ for the reference set and study the effect of variations in two
parameters ($q$ and $\lambda $) and four initial conditions of the scalar
field $f\left( t\right) $ ($f\left( 0\right) $, $f^{\prime }\left( 0\right) $%
, $f^{\prime \prime }\left( 0\right) $, and $f^{\left( 3\right) }\left(
0\right) $) on these quantities. The following is the sets of two parameters
($q$ and $\lambda $) and four initial conditions of the scalar field $%
f\left( t\right) $ ($f\left( 0\right) $, $f^{\prime }\left( 0\right) $, $%
f^{\prime \prime }\left( 0\right) $, and $f^{\left( 3\right) }\left(
0\right) $) used in this numerical computation

\subsection{Results}

The effect of changes in two parameters ($q$ and $\lambda $) and four
initial conditions on the scalar field $f\left( t\right) $ ($f\left(
0\right) $, $f^{\prime }\left( 0\right) $, $f^{\prime \prime }\left(
0\right) $, and $f^{\left( 3\right) }\left( 0\right) $) are studied. Each
parameter/initial condition is varied with respect to the corresponding
starting value. The results are plotted for the scalar field $f\left(
t\right) $, Hubble parameter $H\left( t\right) $, energy density $\rho
\left( t\right) $, pressure $P\left( t\right) $, and the equation of state
(pressure-to-energy density ratio) $w\left( t\right) $. The unspecified
values are the same as that of the reference set Eq \ref{ref}. Here are our results.

\begin{center}
\subsection{Effect of Changes in $q$}
Red $(\cdots)$ \,\,$q=-2$ , Green $(^{\underline{\,\,\,\,\,\,\,}})$\ $q=-6$\,,\,\,Blue (- -)\ $q=-8$
\begin{figure}[hb]
	\centering
		\includegraphics[scale=.3,angle=90]{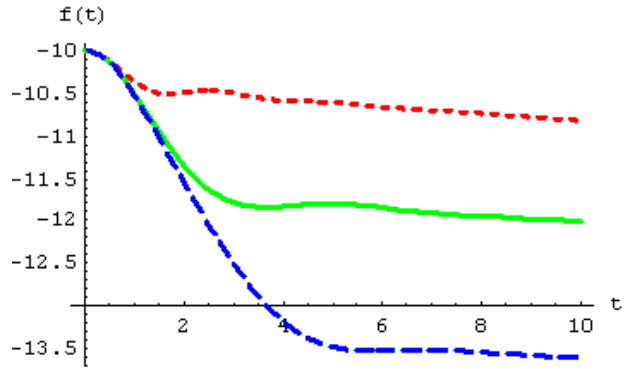}
	\caption{Scalar field $f(t)$ obtained with $q=-2$ (red), $-6$ (green), $-8$ (blue)}
	\label{fig:ff01}
\end{figure}
%
\begin{figure}
	\centering
		\includegraphics[scale=.3,angle=90]{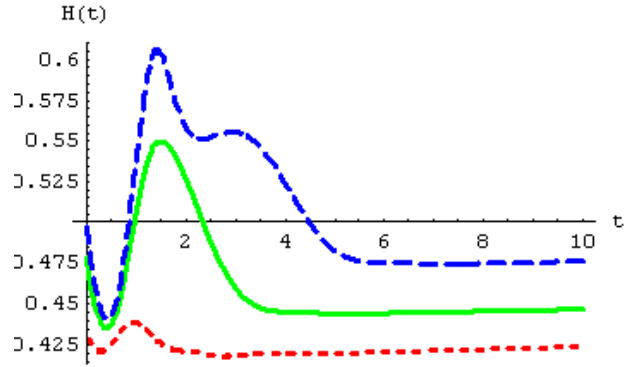}
	\caption{Hubble parameter $H(t)$ obtained with $q=-2$ (red), $-6$ (green), $-8$ (blue)}
	\label{fig:hh01}
\end{figure}
\begin{figure}
	\centering
		\includegraphics[scale=0.3,angle=90]{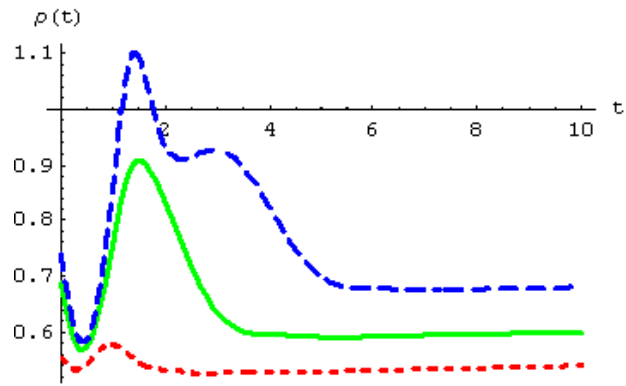}
	\caption{Energy density $\rho(t)$ with $q=-2$ (red), $-6$ (green), $-8$ (blue)}
	\label{fig:rr01}
\end{figure}
\begin{figure}
	\centering
		\includegraphics[scale=0.3,angle=90]{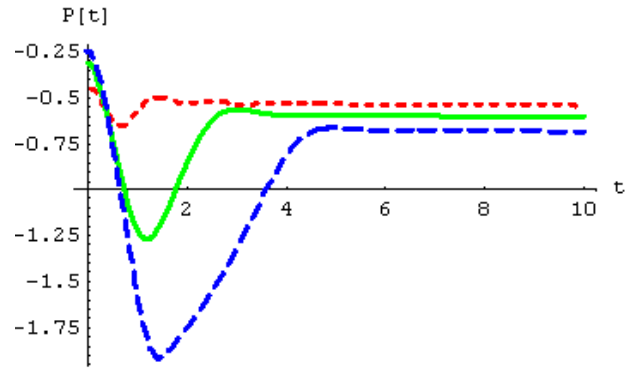}
	\caption{Pressure $p(t)$ obtained with $q=-2$ (red), $-6$ (green), $-8$ (blue)}
	\label{fig:pp01}
\end{figure}
%
\begin{figure}
	\centering
		\includegraphics[scale=0.3,angle=90]{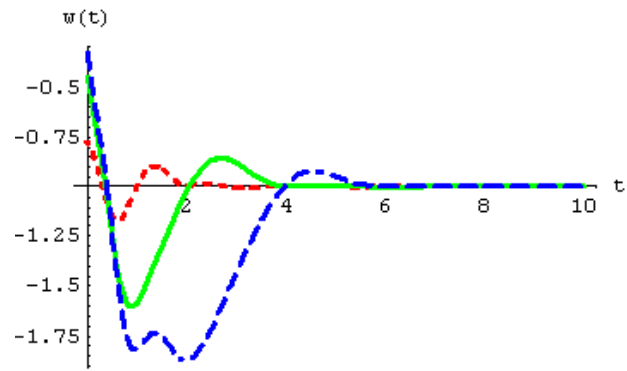}
	\caption{Equation of State $w(t)$ obtained with $q=-2$ (red), $-6$ (green), $-8$ (blue)}
	\label{fig:ww01}
\end{figure}
\begin{figure}
	\centering
		\includegraphics[scale=0.3,angle=90]{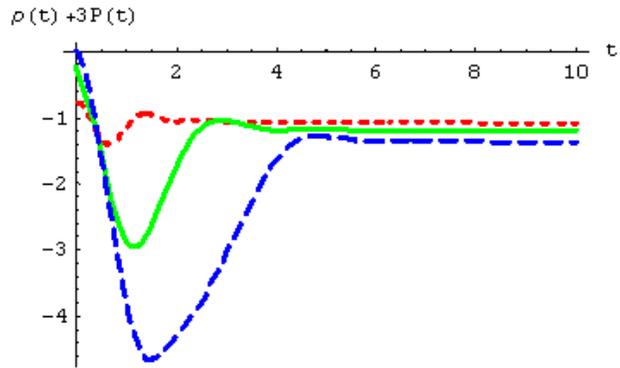}
	\caption{$\rho(t)+3 p(t)$ obtained with $q=-2$ (red), $-6$ (green), $-8$ (blue)}
	\label{fig:rp01}
\end{figure}
\end{center}
\begin{center}
\subsection{Effect of Change in $\lambda $}
Red $(\cdots)$ \,\,$\l=0.08$ , Green $(^{\underline{\,\,\,\,\,\,\,}})$\ $\l=0.05$\,,\,\,Blue (- -)\ $\l=0.02$
\begin{figure}[hb]
	\centering
		\includegraphics[scale=0.30,angle=90]{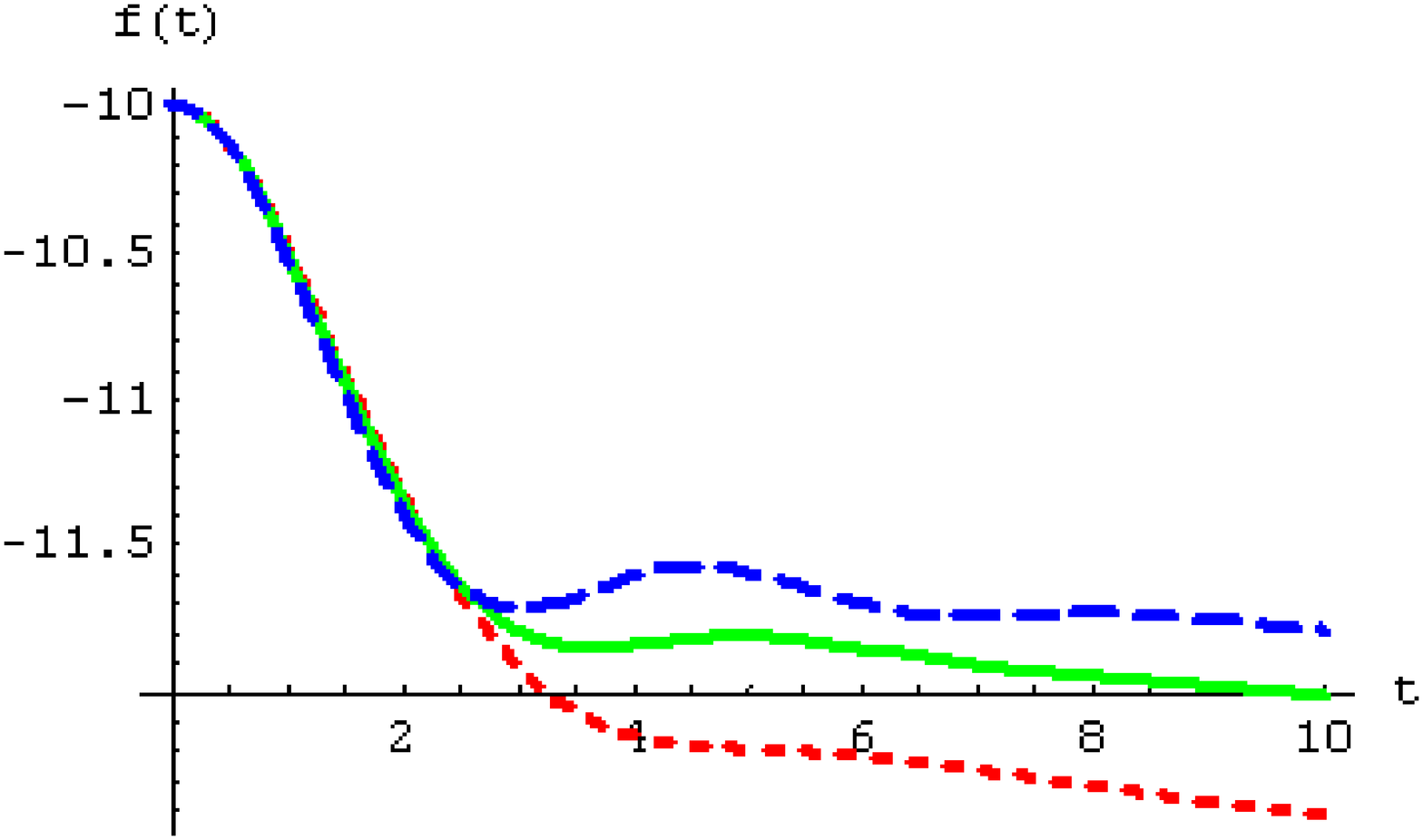}
	\caption{$f(t)$ obtained with $\l=0.08$ (red), $0.05$ (green), $0.02$ (blue)}
	\label{fig:ff02}
\end{figure}
%
\begin{figure}
	\centering
		\includegraphics[scale=0.30,angle=90]{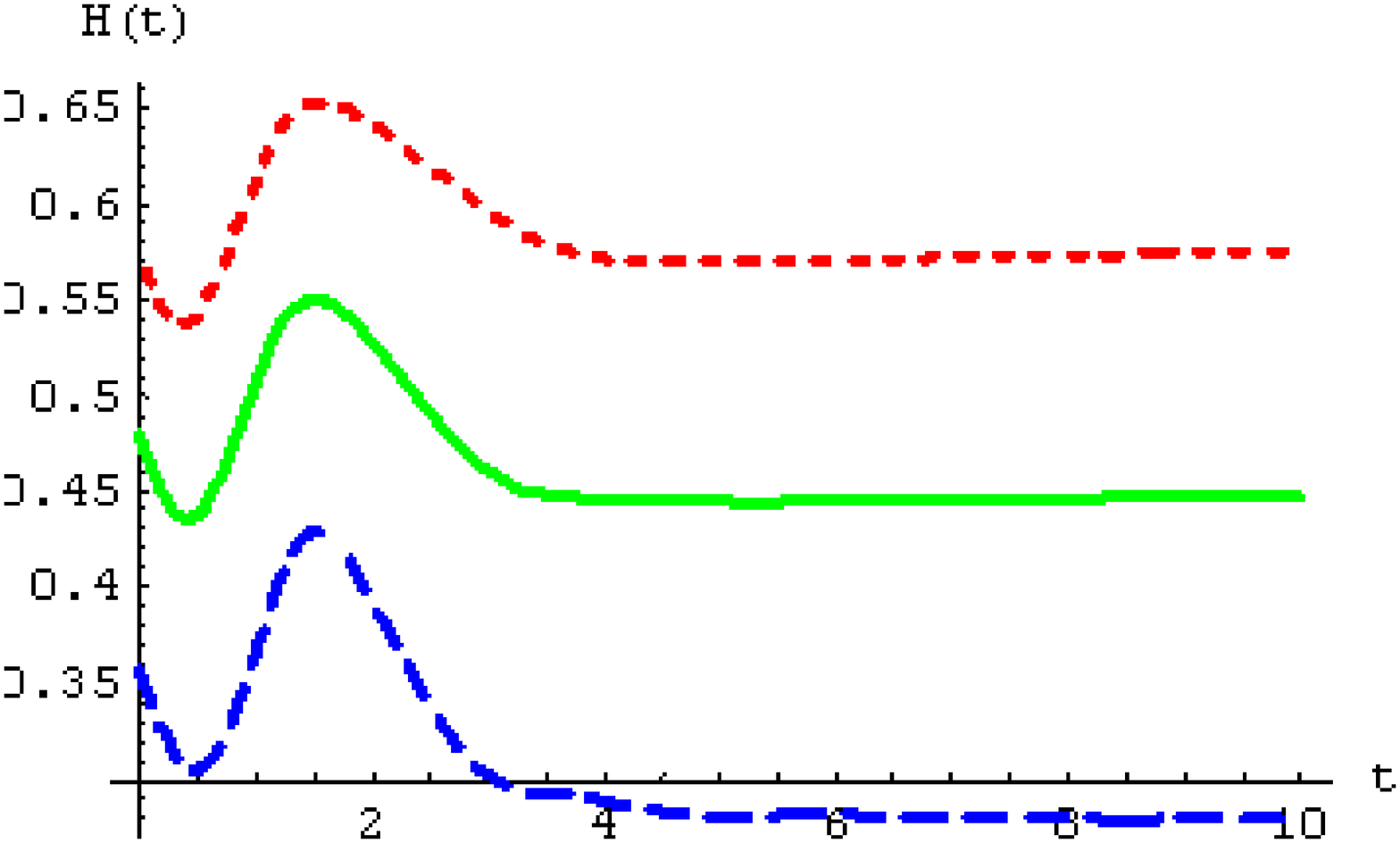}
	\caption{$H(t)$ obtained with $\l=0.08$ (red), $0.05$ (green), $0.02$ (blue)}
	\label{fig:hh02}
\end{figure}
%
\begin{figure}
	\centering
		\includegraphics[scale=0.3,angle=90]{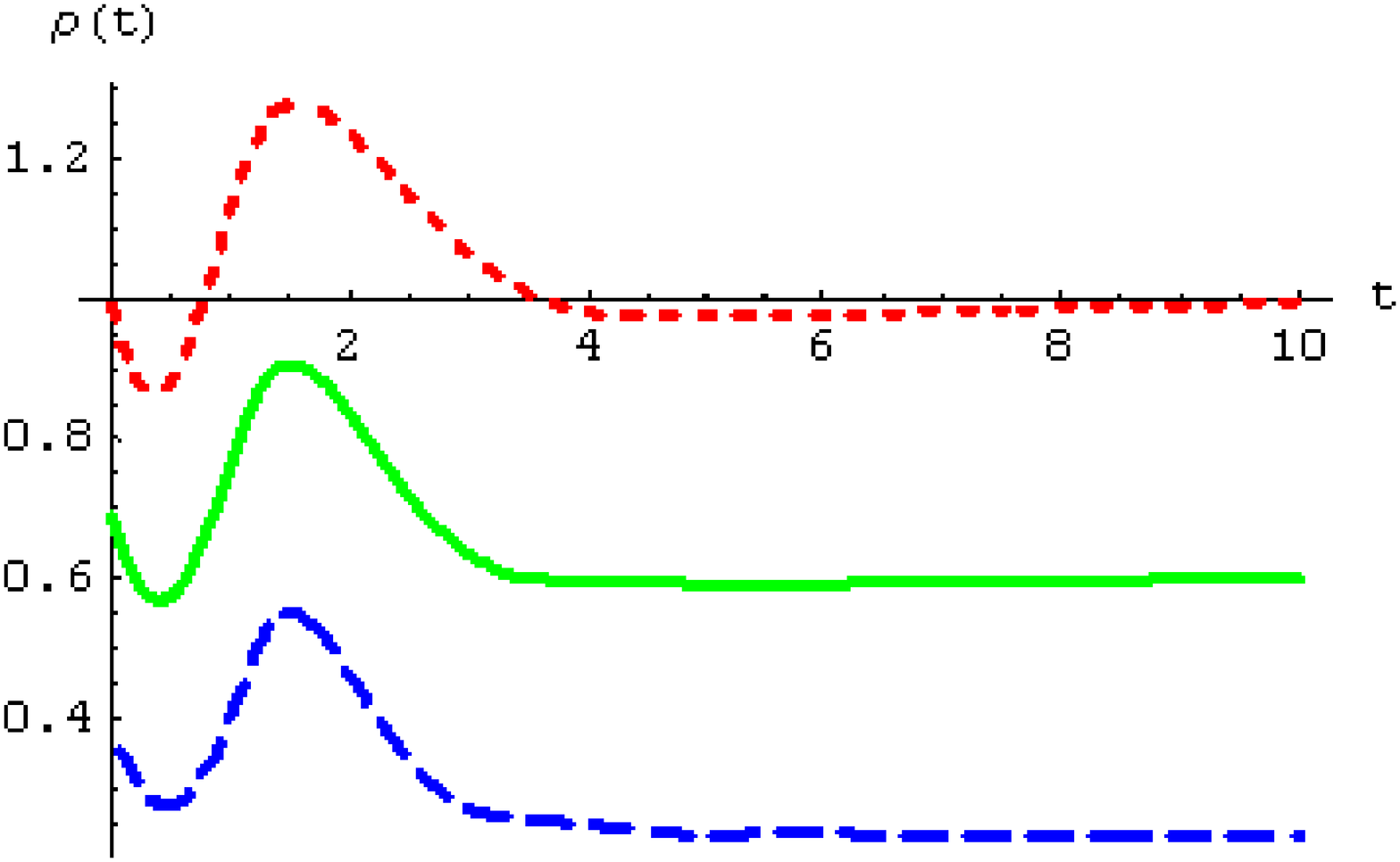}
	\caption{$\r(t)$ obtained with $\l=0.08$ (red), $0.05$ (green), $0.02$ (blue)}
	\label{fig:rr02}
\end{figure}
\begin{figure}
	\centering
		\includegraphics[scale=0.3,angle=90]{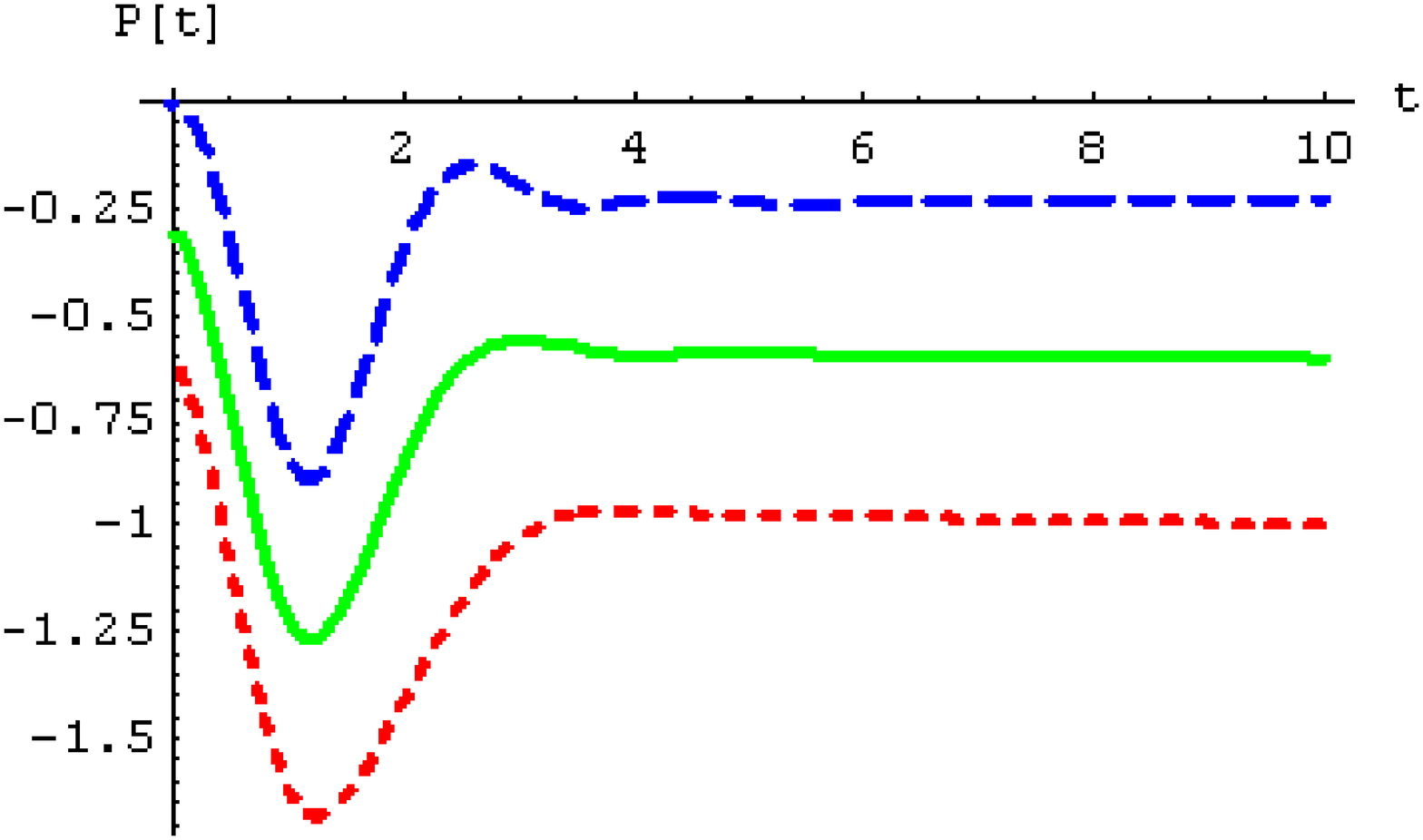}
	\caption{$p(t)$ obtained with $\l=0.08$ (red), $0.05$ (green), $0.02$ (blue)}
	\label{fig:pp02}
\end{figure}
%
\begin{figure}
	\centering
		\includegraphics[scale=0.3,angle=90]{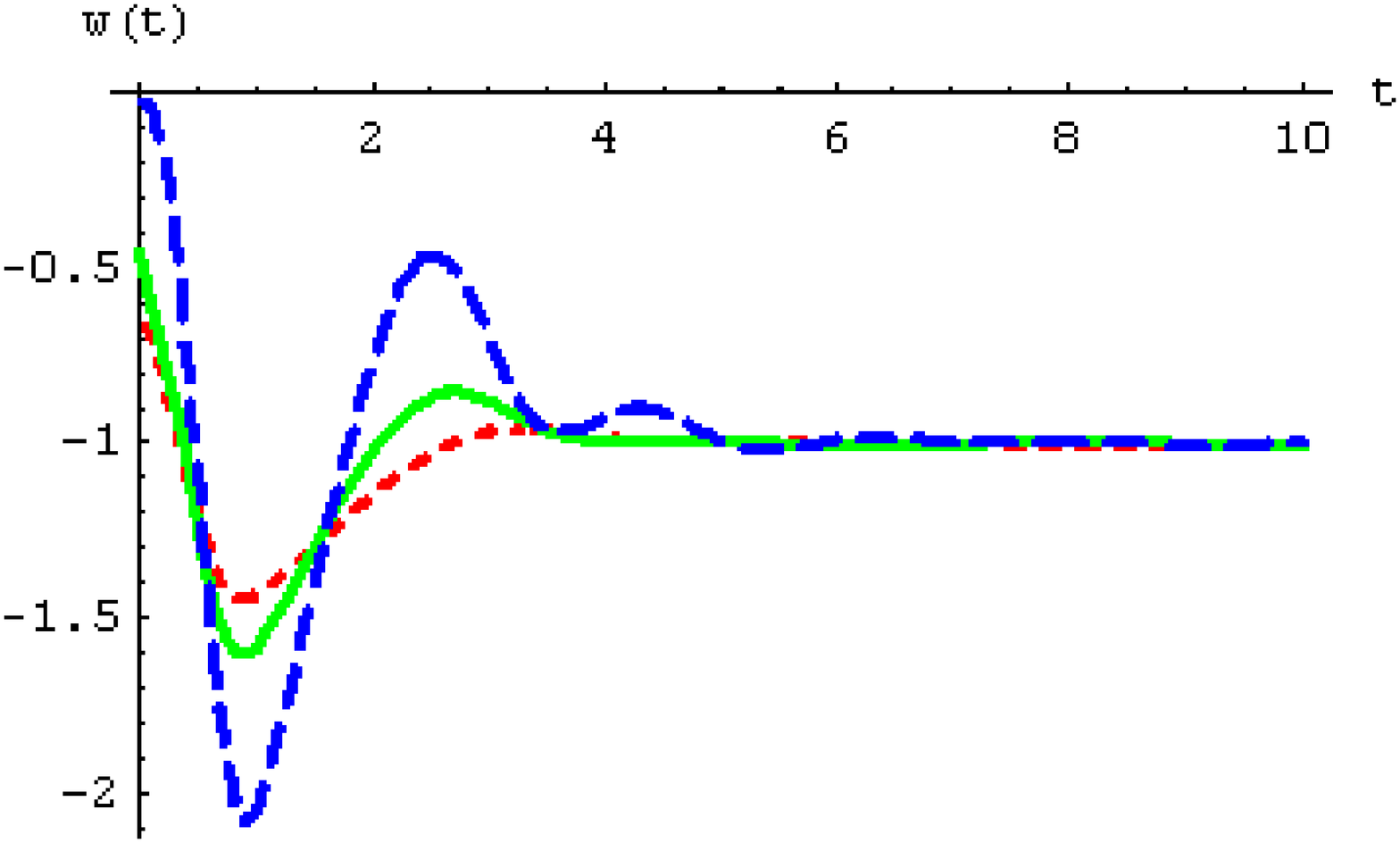}
	\caption{Equation of State $w(t)$ obtained with $\l=0.08$ (red), $0.05$ (green), $0.02$ (blue)}
	\label{fig:ww02}
\end{figure}
%
\begin{figure}
	\centering
		\includegraphics[scale=0.3,angle=90]{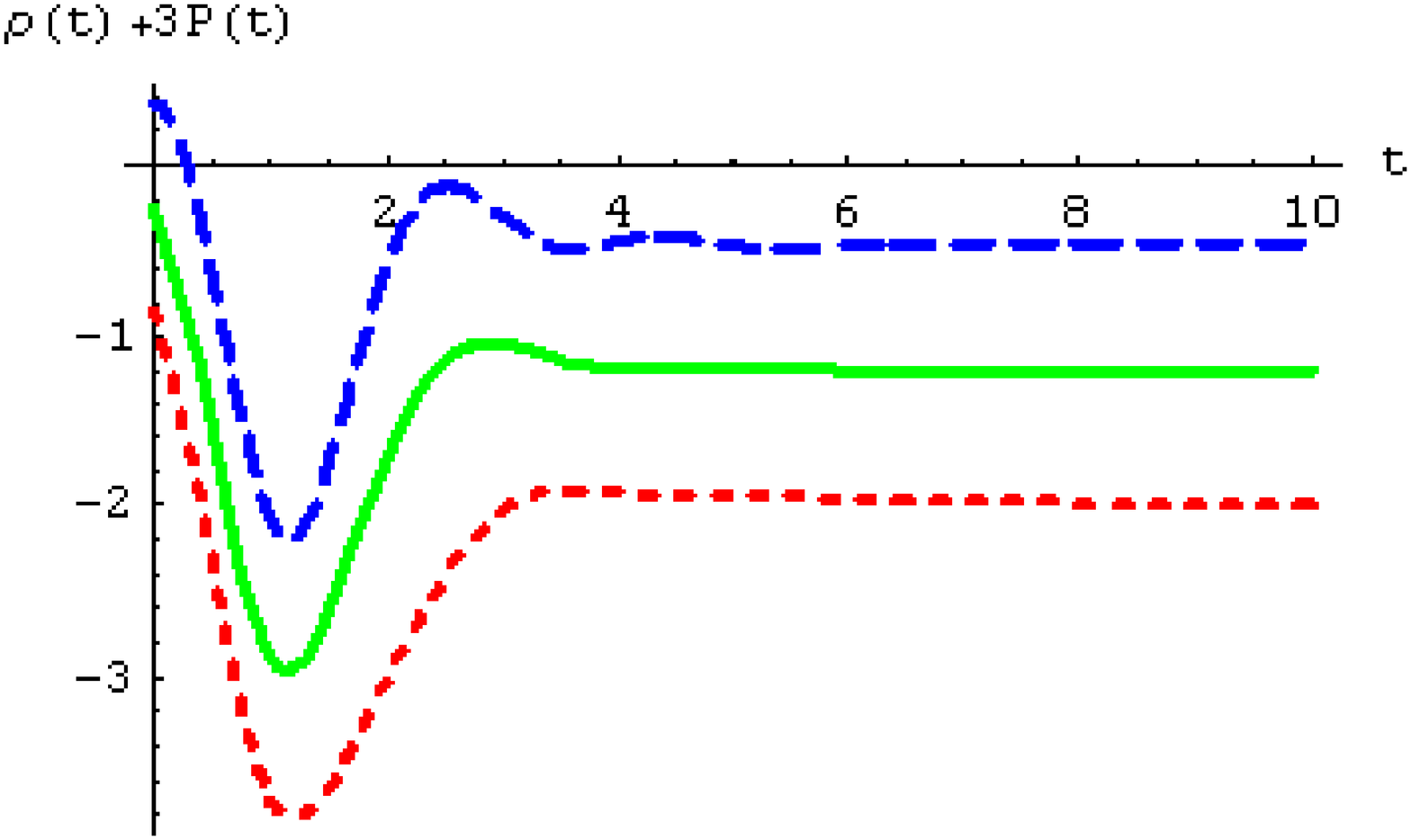}
	\caption{$\r(t)+3p(t)$ obtained with $\l=0.08$ (red), $0.05$ (green), $0.02$ (blue)}
	\label{fig:rp02}
\end{figure}
\end{center}

\pagebreak

\begin{center}
\subsection{Effect of Change in Initial Value of $f\left( t_{0}\right) $}%
Red $(\cdots)$ \,\,$f(t_0)=-8$ , Green $(^{\underline{\,\,\,\,\,\,\,}})$\ $f(t_0)=-10$\,,\,\,Blue (- -)\ $f(t_0)=-12$
\begin{figure}[hb]
	\centering
		\includegraphics[scale=0.3,angle=90]{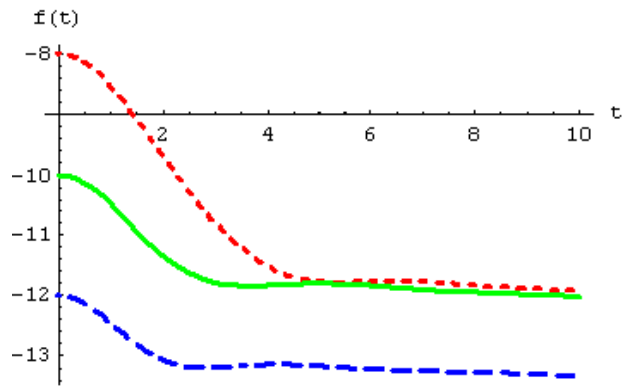}
	\caption{Scalar field $f(t)$ obtained with $f(t_0)=-8$ (red), $-10$ (green), $-12$ (blue)}
	\label{fig:ff03}
\end{figure}
\begin{figure}
	\centering
		\includegraphics[scale=0.3,angle=90]{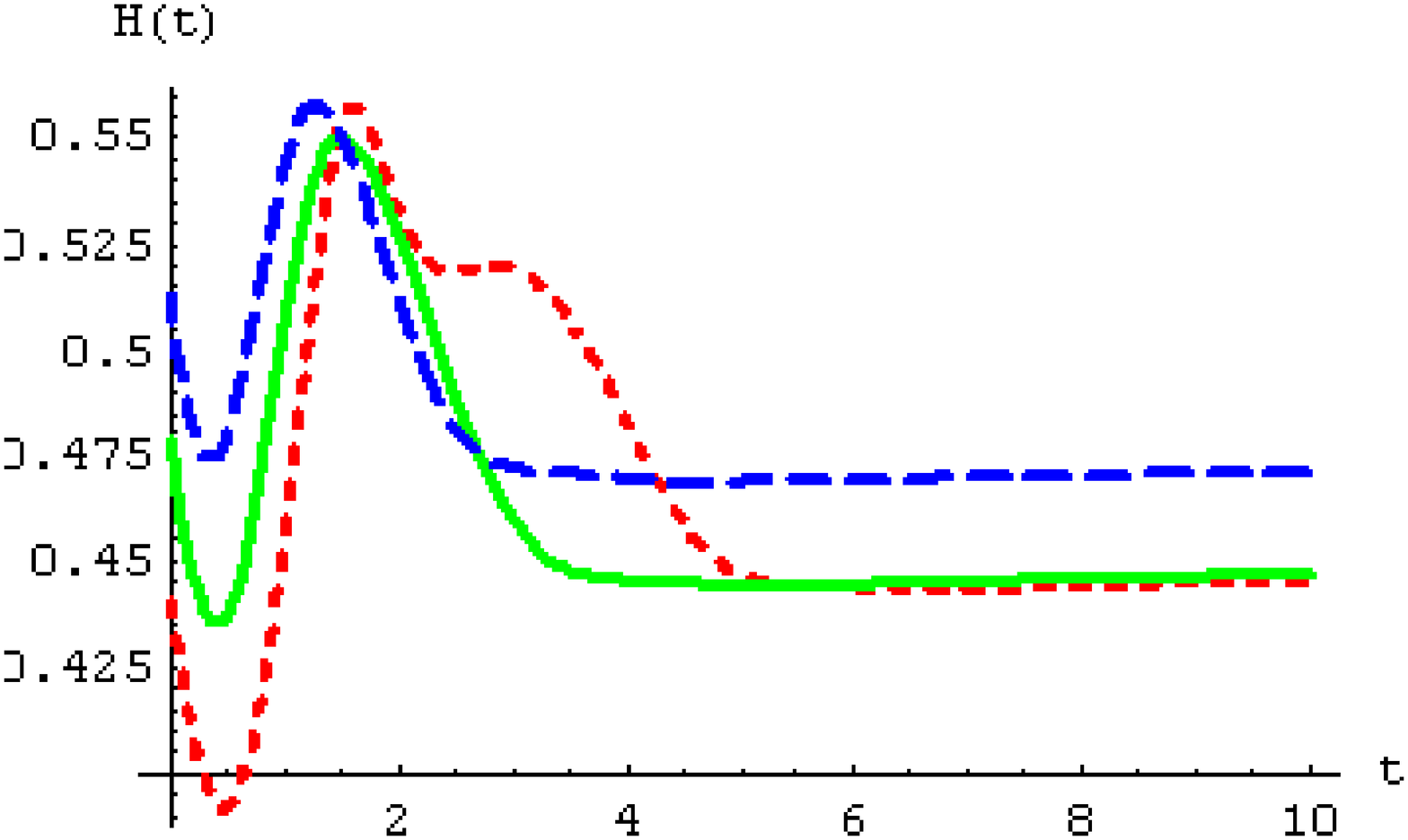}
	\caption{Hubble parameter $H(t)$ obtained with $f(t_0)=-8$ (red), $-10$ (green), $-12$ (blue)}
	\label{fig:hh03}
\end{figure}
\begin{figure}
	\centering
		\includegraphics[scale=0.3,angle=90]{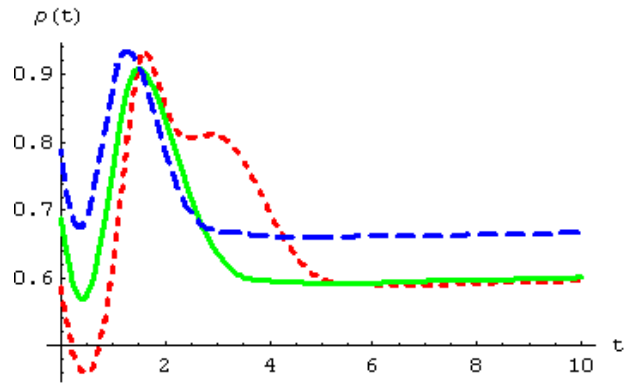}
	\caption{Energy Density $\r(t)$ obtained with $f(t_0)=-8$ (red), $-10$ (green), $-12$ (blue)}
	\label{fig:rr03}
\end{figure}
%
\begin{figure}
	\centering
		\includegraphics[scale=0.3,angle=90]{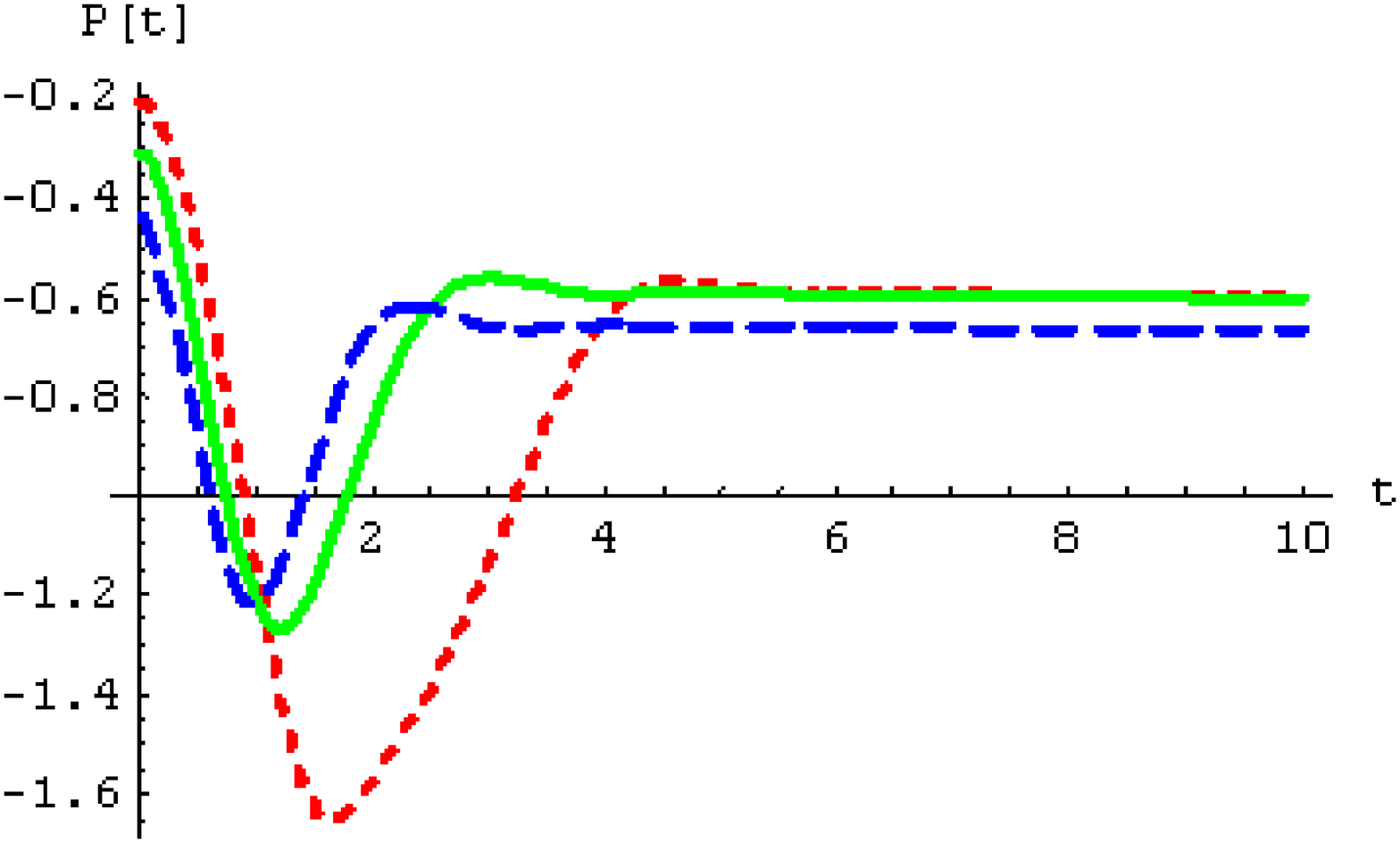}
	\caption{Pressure $p(t)$ obtained with $f(t_0)=-8$ (red), $-10$ (green), $-12$ (blue)}
	\label{fig:pp03}
\end{figure}
%
\begin{figure}
	\centering
		\includegraphics[scale=0.3,angle=90]{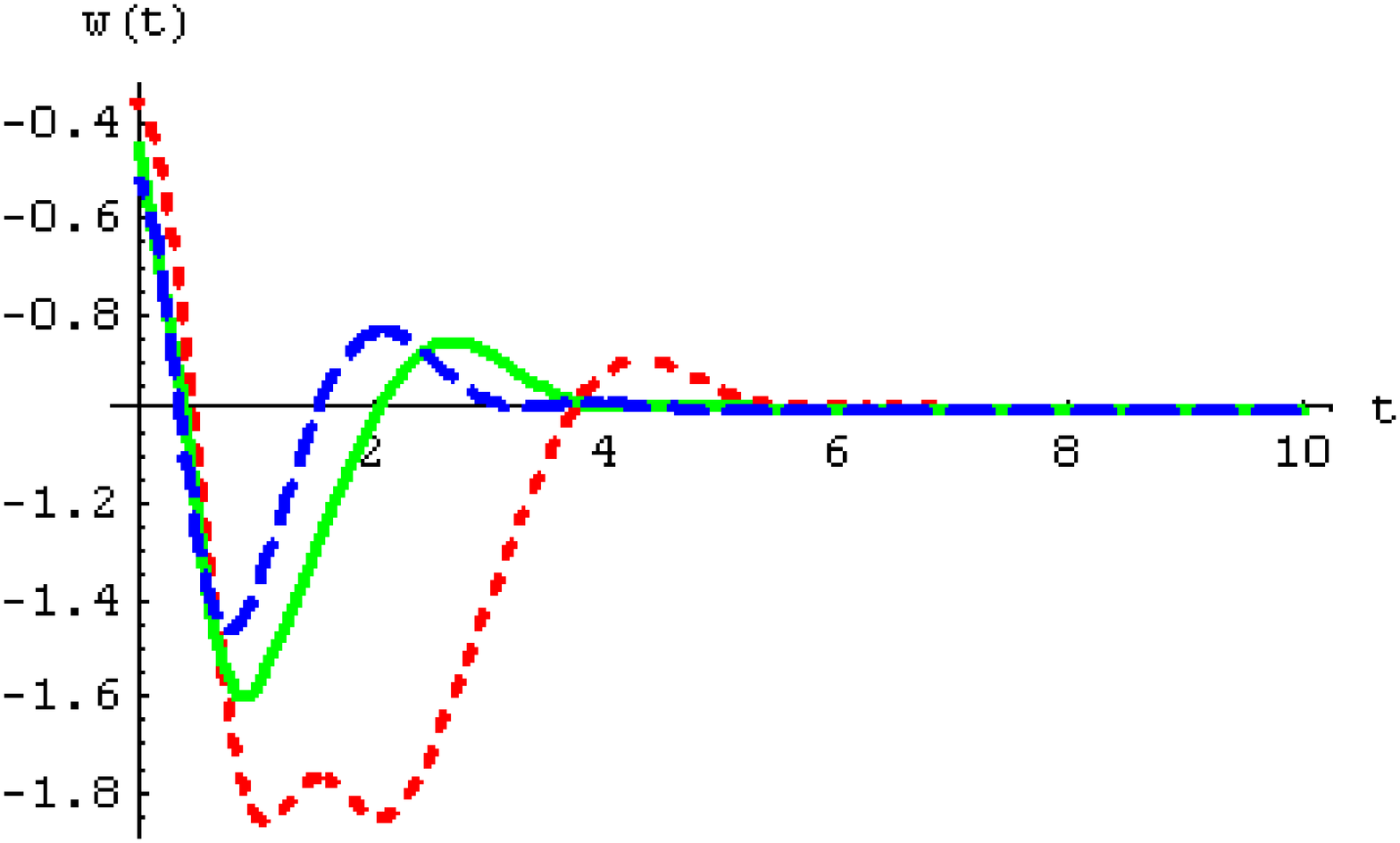}
	\caption{Equation of State $w(t)$ obtained with $f(t_0)=-8$ (red), $-10$ (green), $-12$ (blue)}
	\label{fig:ww03}
\end{figure}
%
\begin{figure}
	\centering
		\includegraphics[scale=0.3,angle=90]{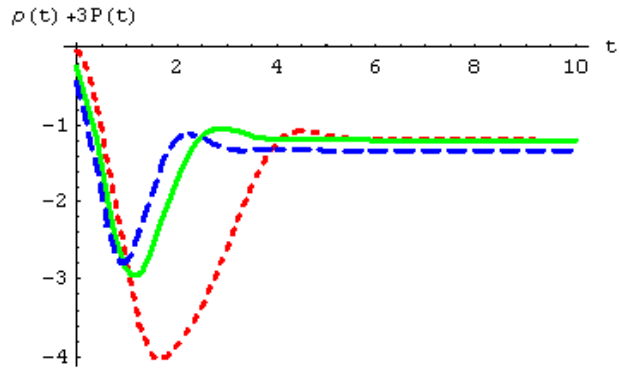}
	\caption{$\r(t)+3 p(t)$ obtained with $f(t_0)=-8$ (red), $-10$ (green), $-12$ (blue)}
	\label{fig:rp03}
\end{figure}
\pagebreak
\subsection{Effect of Change in Initial Value of $f^{\prime }\left( t_{0}\right) $}
Red $(\cdots)$ \,\,$f^{\prime }\left( t_{0}\right)=-0.5$ , Green $(^{\underline{\,\,\,\,\,\,\,}})$\ $f^{\prime }\left( t_{0}\right)=0$\,,\,\,Blue (- -)\ $f^{\prime }\left( t_{0}\right)=0.5$
\begin{figure}[hb]
	\centering
		\includegraphics[scale=.3,angle=90]{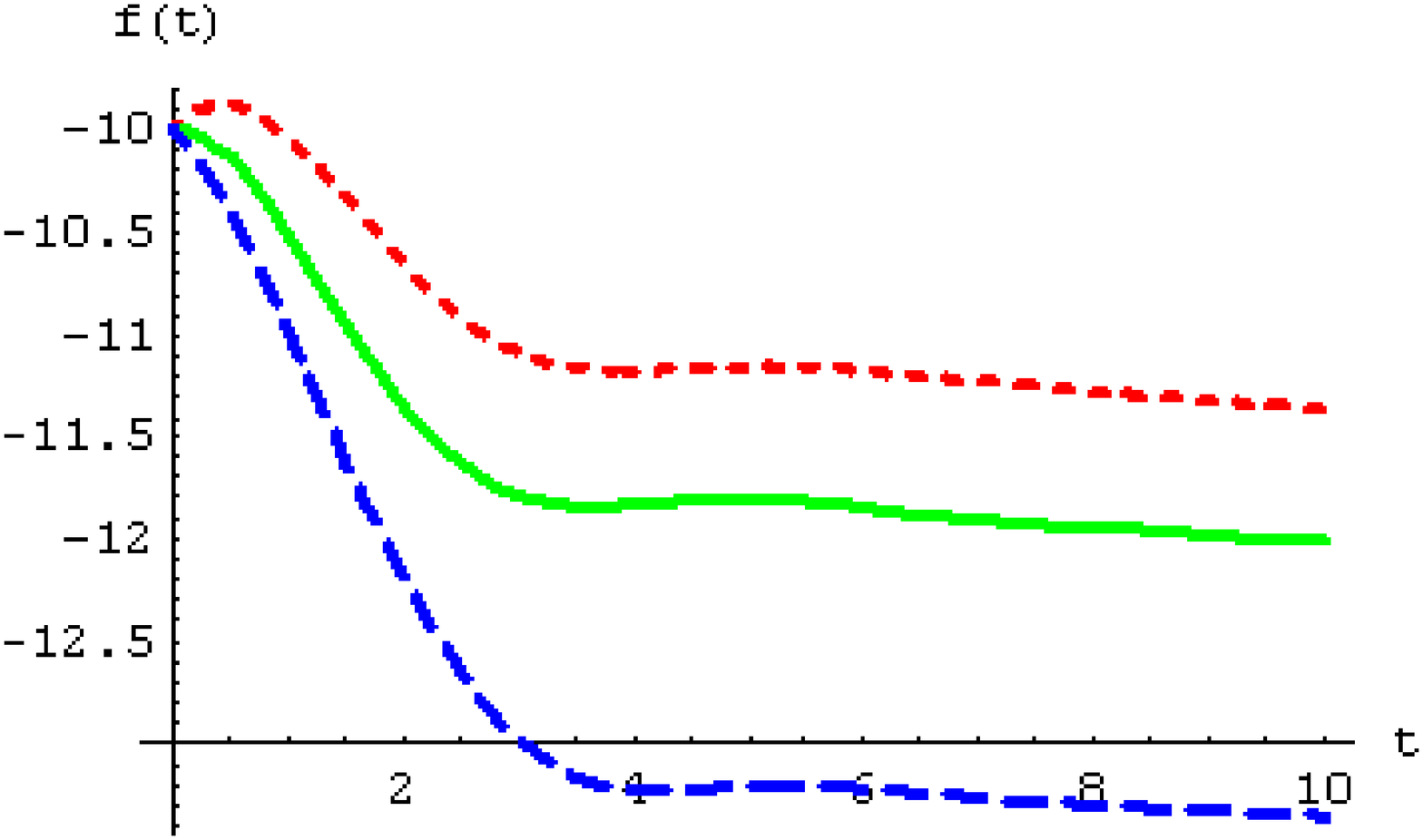}
	\caption{Scalar field $f(t)$ obtained with $f^{\prime}(t_0)=-0.52$ (red), $0$ (green), $0.5$ (blue)}
	\label{fig:ff04}
\end{figure}
\begin{figure}
	\centering
		\includegraphics[scale=.3,angle=90]{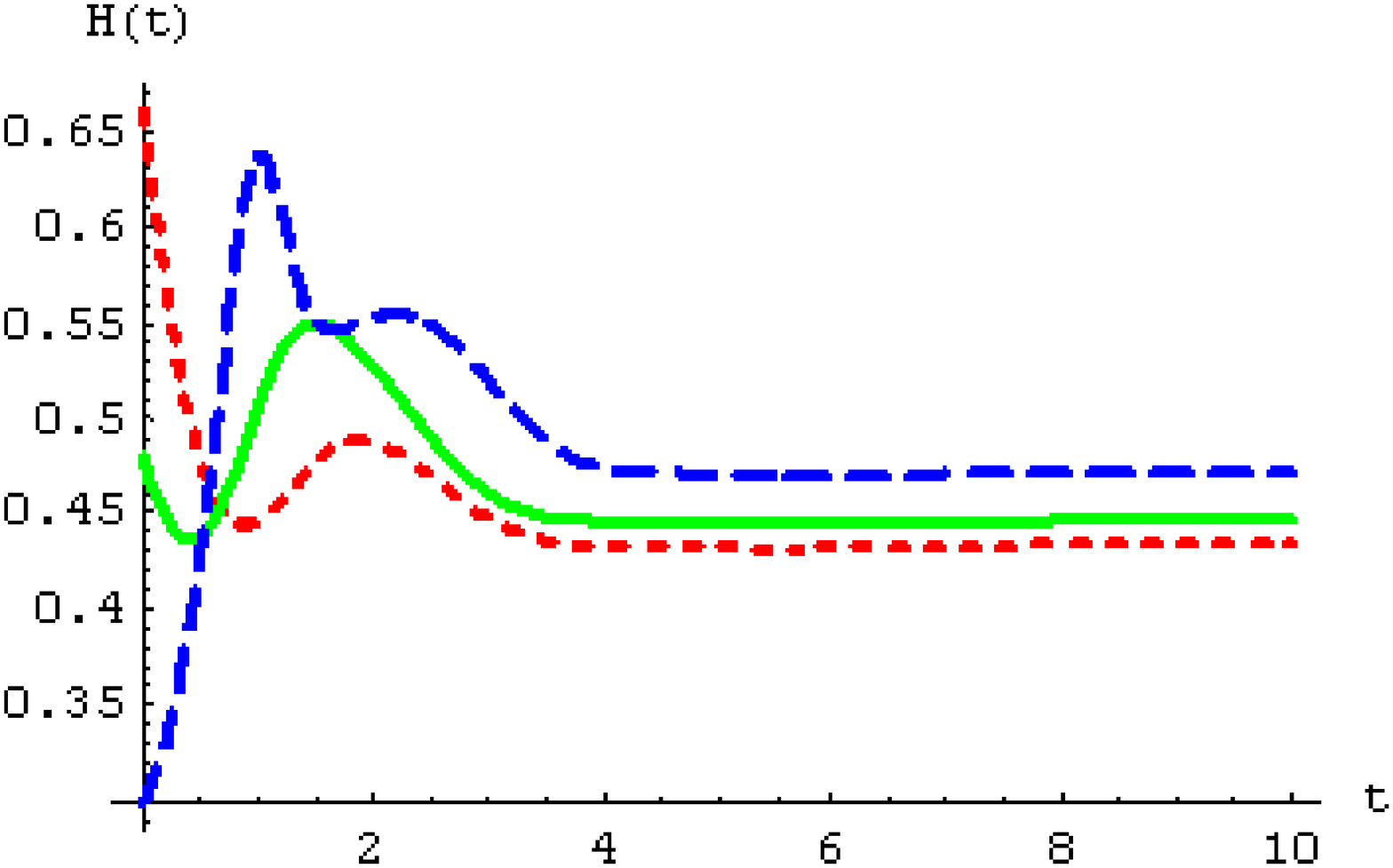}
	\caption{Scalar field $H(t)$ obtained with $f^{\prime}(t_0)=-0.52$ (red), $0$ (green), $0.5$ (blue)}
	\label{fig:hh04}
\end{figure}
%
\begin{figure}
	\centering
		\includegraphics[scale=.3,angle=90]{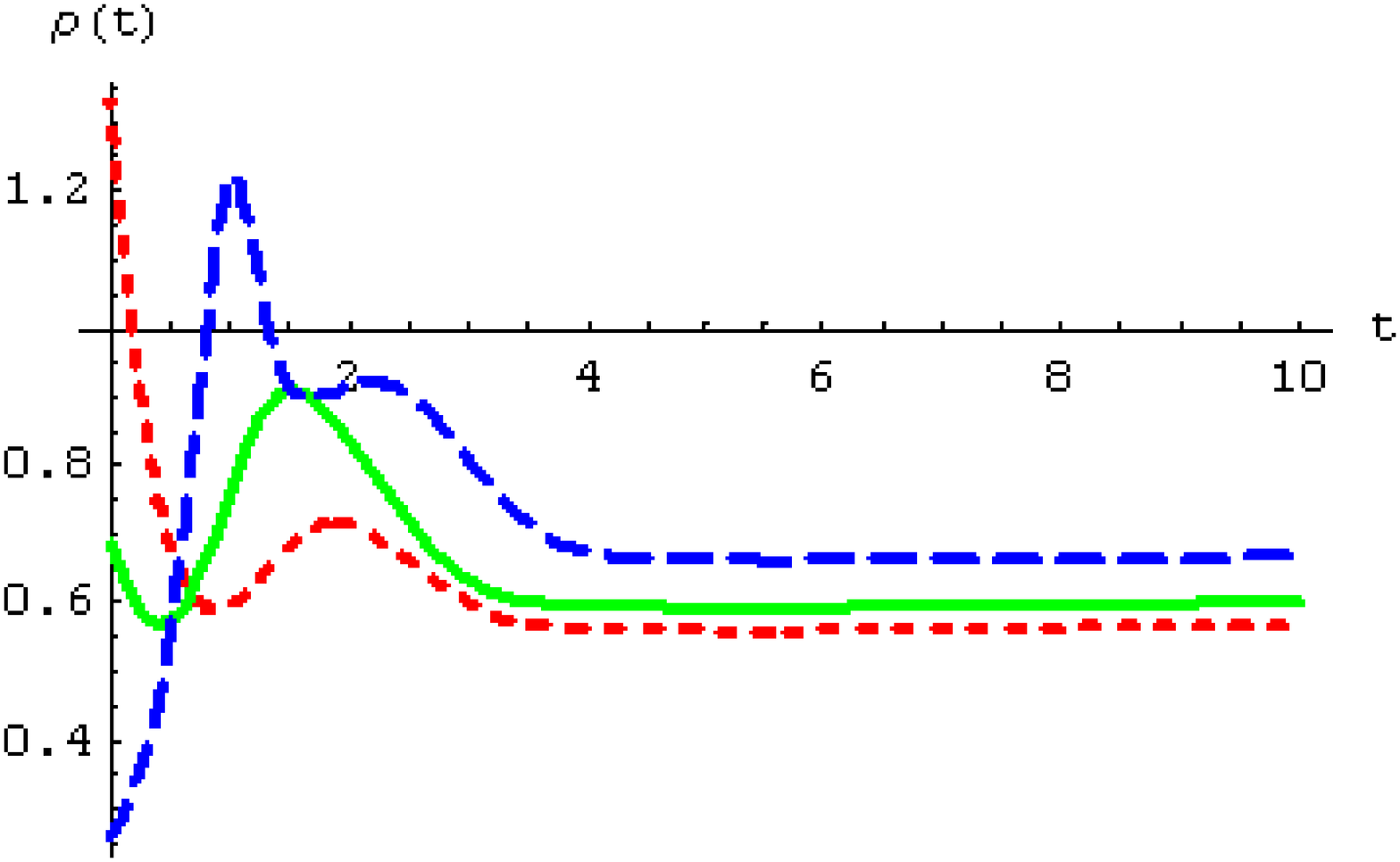}
	\caption{Energy density $\r(t)$ obtained with $f^{\prime}(t_0)=-0.52$ (red), $0$ (green), $0.5$ (blue)}
	\label{fig:rr04}
\end{figure}
\begin{figure}
	\centering
		\includegraphics[scale=.3,angle=90]{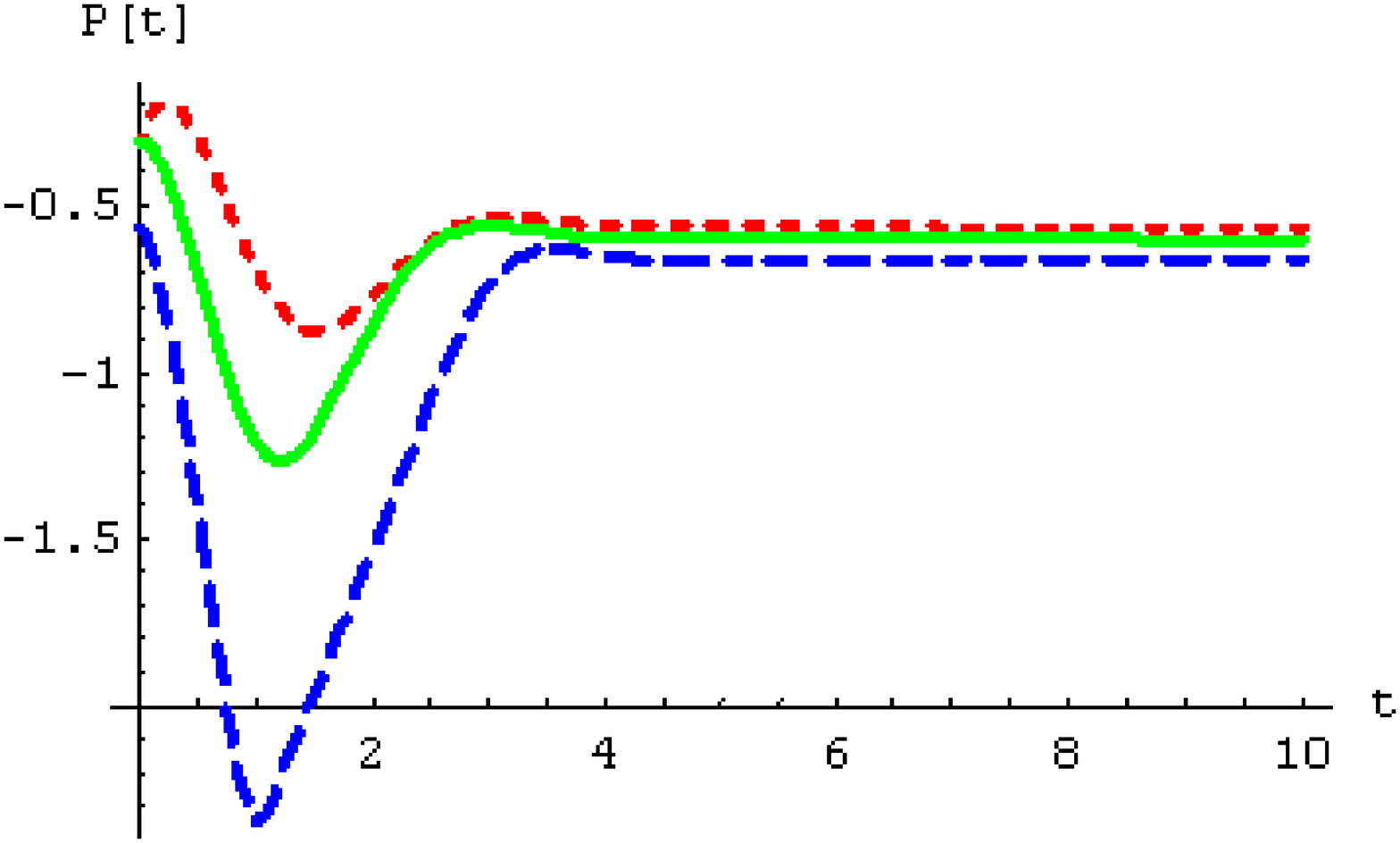}
	\caption{Pressure $p(t)$ obtained with $f^{\prime}(t_0)=-0.52$ (red), $0$ (green), $0.5$ (blue)}
	\label{fig:pp04}
\end{figure}
%
\begin{figure}
	\centering
		\includegraphics[scale=.3,angle=90]{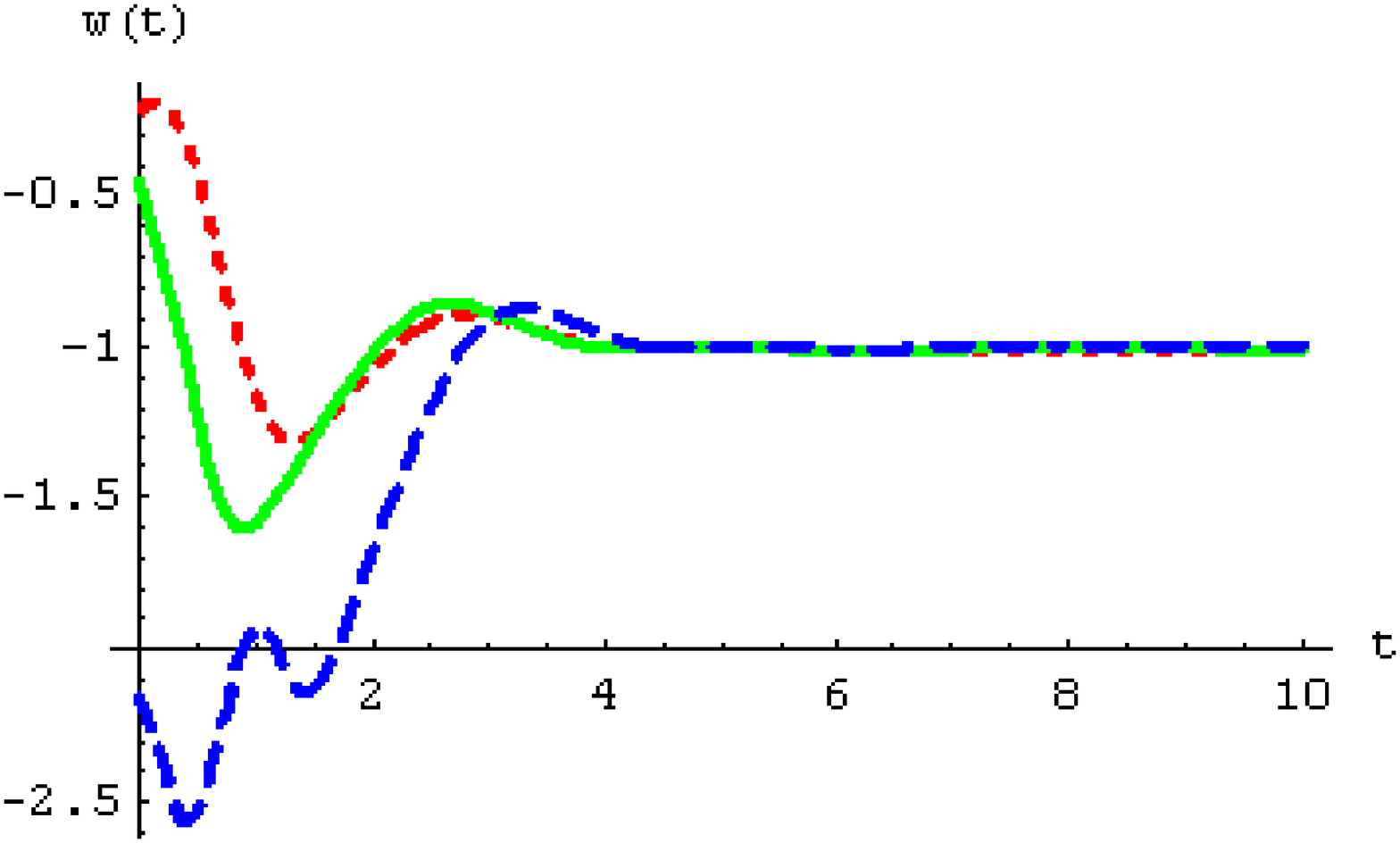}
	\caption{Equation of state $w(t)$ obtained with $f^{\prime}(t_0)=-0.52$ (red), $0$ (green), $0.5$ (blue)}
	\label{fig:ww04}
\end{figure}
%
\begin{figure}
	\centering
		\includegraphics[scale=.3,angle=90]{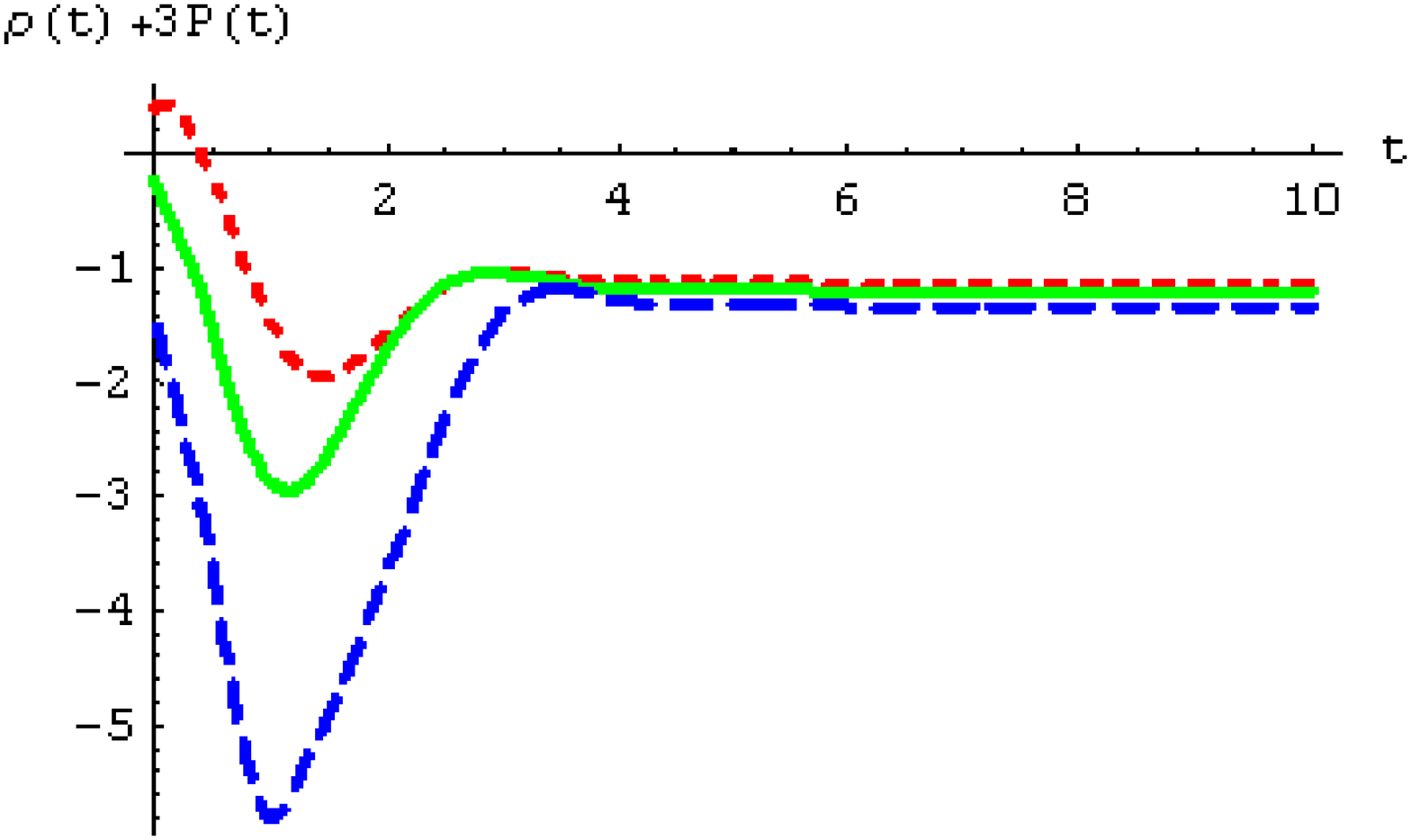}
	\caption{$\r(t)+3 p(t)$ obtained with $f^{\prime}(t_0)=-0.52$ (red), $0$ (green), $0.5$ (blue)}
	\label{fig:rp04}
\end{figure}
\pagebreak
\subsection{Effect of Change in Initial Value of $f^{\prime \prime }\left( t_{0}\right) $}
Red $(\cdots)$ \,\,$f^{\prime \prime }(t_0)=-0.5$ , Green $(^{\underline{\,\,\,\,\,\,\,}})$\ $f^{\prime \prime }(t_0)=-1$\,,\,\,Blue (- -)\ $f^{\prime \prime }(t_0)=-1.5$
\begin{figure}[hb]
	\centering
		\includegraphics[scale=0.3,angle=90]{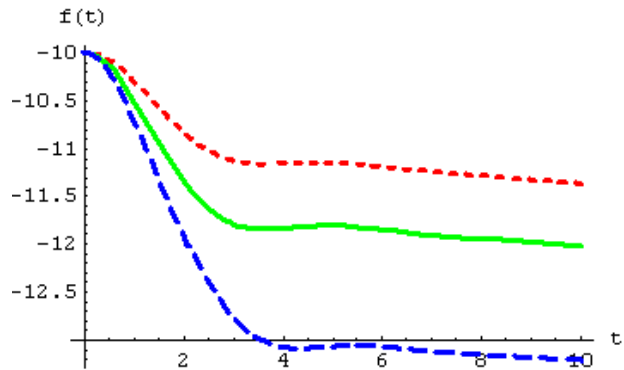}
	\caption{Scalar field $f(t)$  with $f^{\prime \prime}(t_0)=-8$ (red), $-10$ (green), $-12$ (blue)}
	\label{fig:ff05}
\end{figure}
\begin{figure}
	\centering
		\includegraphics[scale=0.3,angle=90]{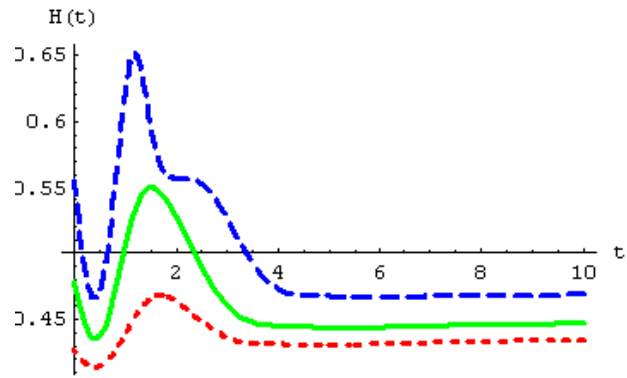}
	\caption{Hubble parameter $H(t)$  with $f^{\prime \prime}(t_0)=-8$ (red), $-10$ (green), $-12$ (blue)}
	\label{fig:hh05}
\end{figure}
%
\begin{figure}
	\centering
		\includegraphics[scale=0.3,angle=90]{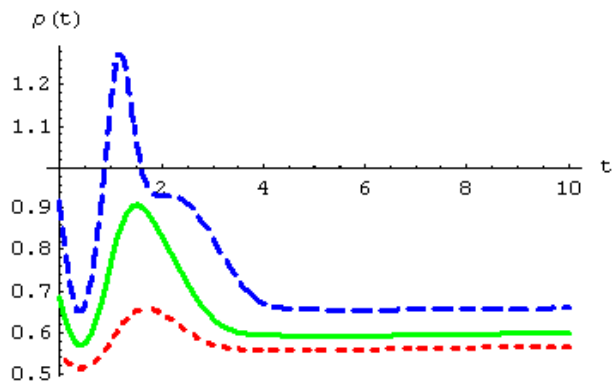}
	\caption{Energy density $\r(t)$ with $f^{\prime \prime}(t_0)=-8$ (red), $-10$ (green), $-12$ (blue)}
	\label{fig:rr05}
\end{figure}
\begin{figure}
	\centering
		\includegraphics[scale=0.3,angle=90]{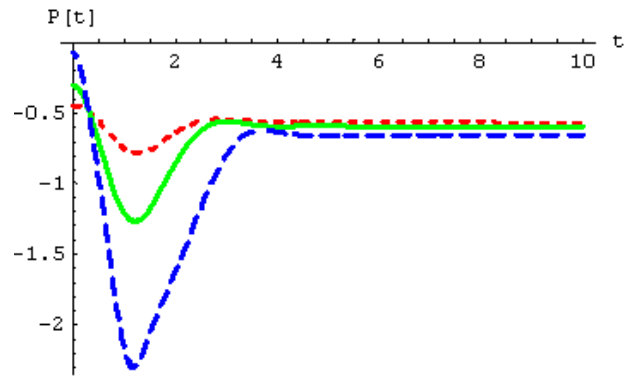}
	\caption{Pressure $p(t)$ with $f^{\prime \prime}(t_0)=-8$ (red), $-10$ (green), $-12$ (blue)}
	\label{fig:pp05}
\end{figure}
%
\begin{figure}
	\centering
		\includegraphics[scale=0.3,angle=90]{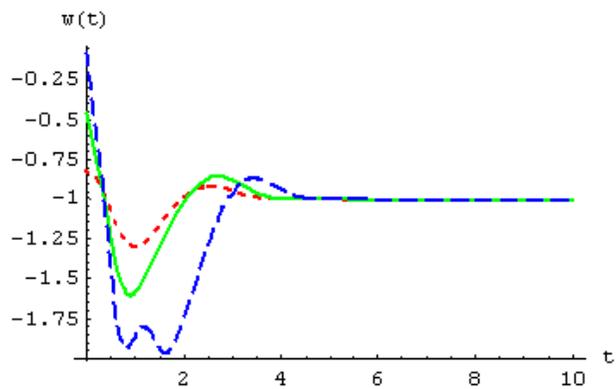}
	\caption{Equation of state $w(t)$ with $f^{\prime \prime}(t_0)=-8$ (red), $-10$ (green), $-12$ (blue)}
	\label{fig:ww05}
\end{figure}
%
\begin{figure}
	\centering
		\includegraphics[scale=0.3,angle=90]{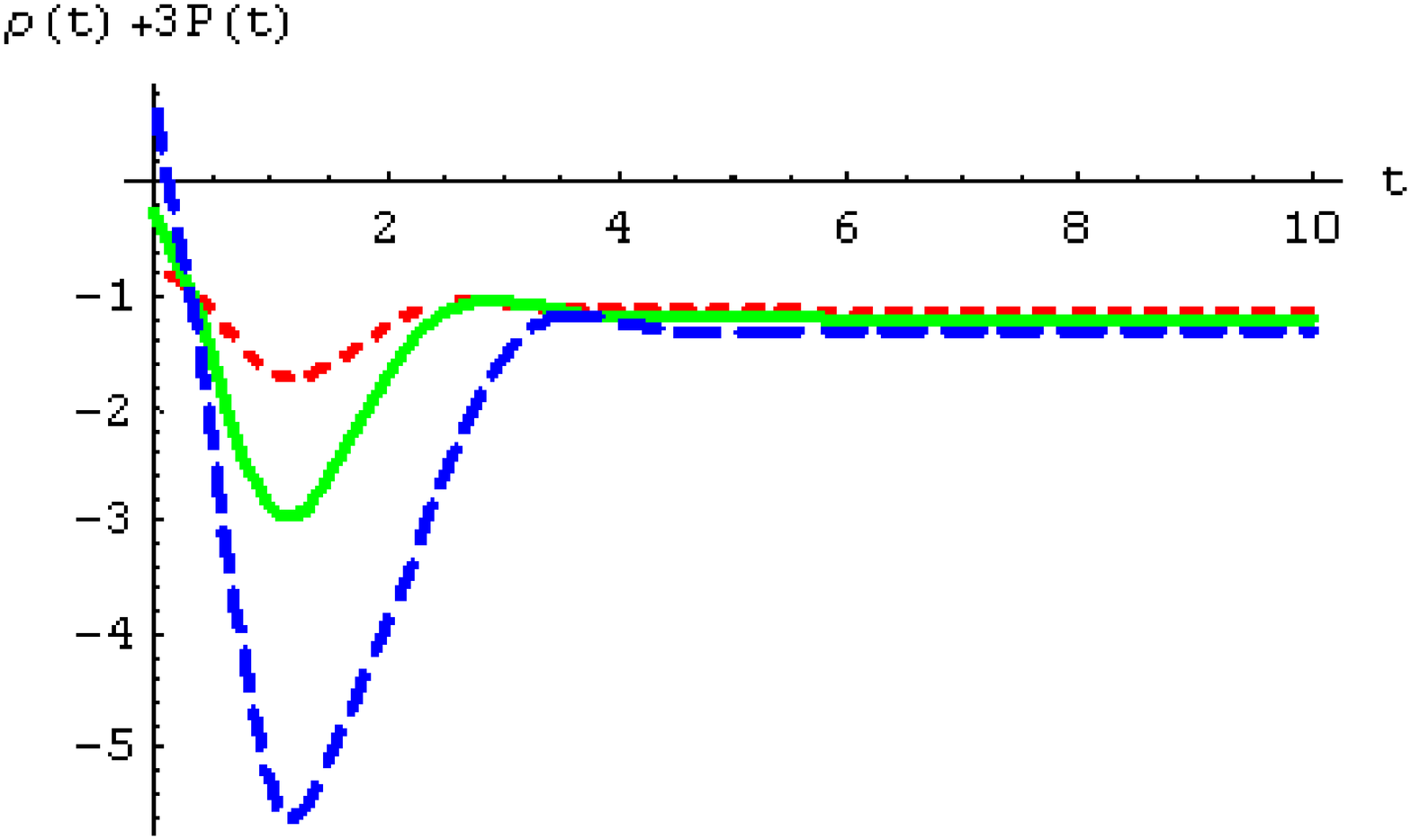}
	\caption{$\r(t)+3 p(t)$ with $f^{\prime \prime}(t_0)=-8$ (red), $-10$ (green), $-12$ (blue)}
	\label{fig:rp05}
\end{figure}

\subsection{Effect of Change in Initial Value of $f^{\left( 3\right) }\left(
t_{0}\right) $}
Red $(\cdots)$ \,\,$f^{\left(3\right)}=-0.5$ , Green $(^{\underline{\,\,\,\,\,\,\,}})$\ $f^{\left(3\right)}=-1$\,,\,\,Blue (- -)\ $f^{\left(3\right)}=-1.5$
\begin{figure}
	\centering
		\includegraphics[scale=0.3,angle=90]{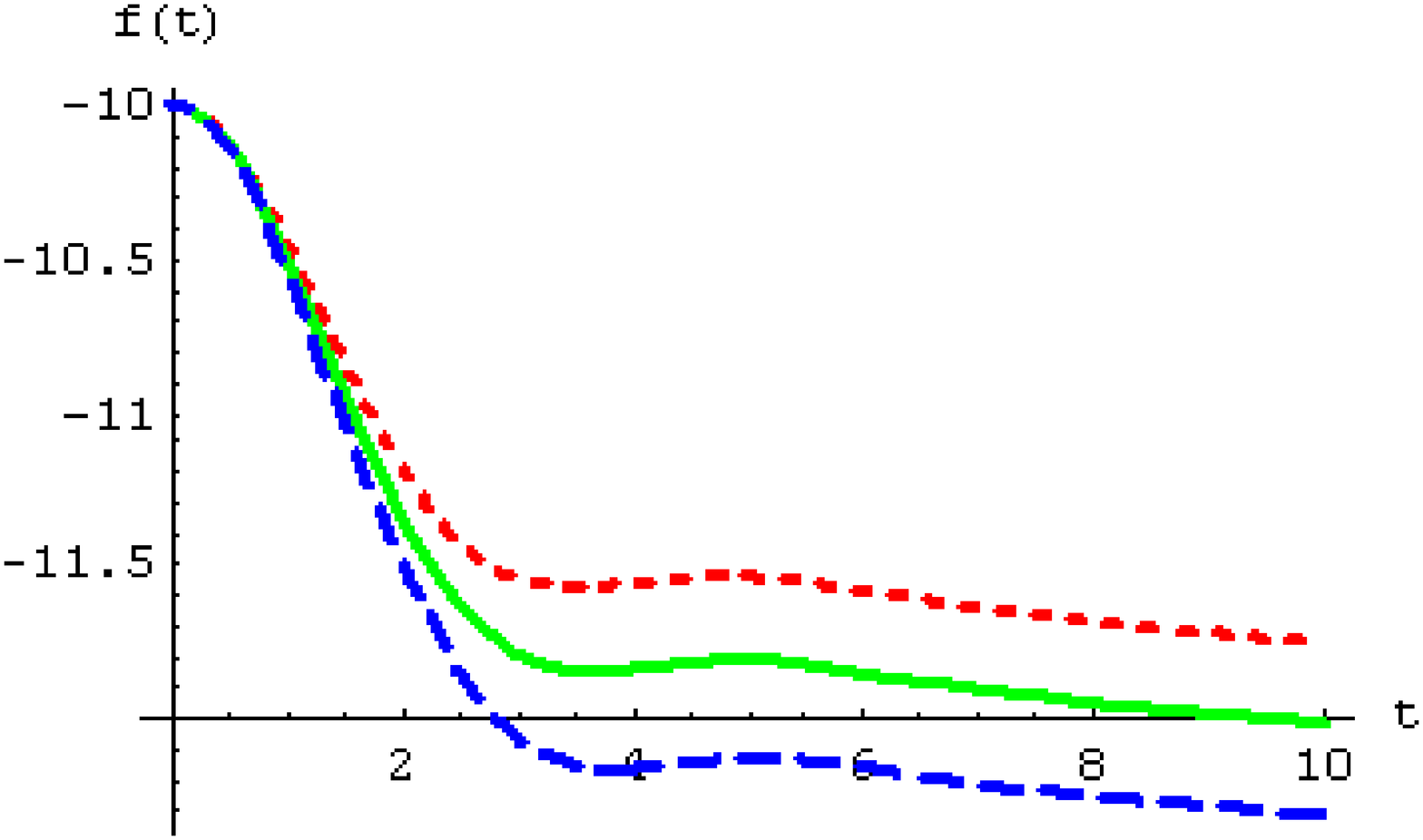}
	\caption{Scalar field $f(t)$ obtained with $f^{(3)}(t_0)=-0.5$ (red), $-1$ (green), $-1.5$ (blue)}
	\label{fig:ff06}
\end{figure}
\begin{figure}
	\centering
		\includegraphics[scale=0.3,angle=90]{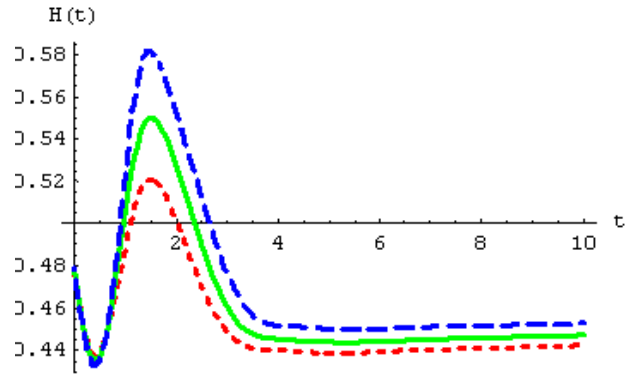}
	\caption{Hubble parameter $H(t)$ obtained with $f^{(3)}(t_0)=-0.5$ (red), $-1$ (green), $-1.5$ (blue)}
	\label{fig:hh06}
\end{figure}
\begin{figure}
	\centering
		\includegraphics[scale=0.3,angle=90]{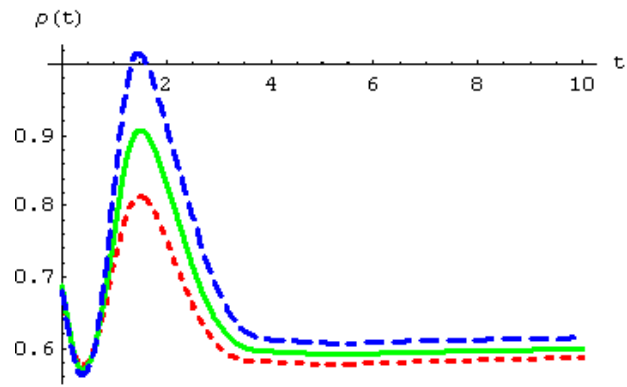}
	\caption{Energy density $\r(t)$ obtained with $f^{(3)}(t_0)=-0.5$ (red), $-1$ (green), $-1.5$ (blue)}
	\label{fig:rr06}
\end{figure}
\begin{figure}
	\centering
		\includegraphics[scale=0.3,angle=90]{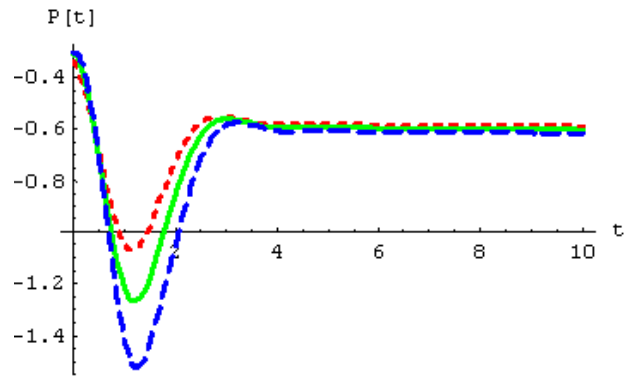}
	\caption{Pressure $p(t)$ obtained with $f^{(3)}(t_0)=-0.5$ (red), $-1$ (green), $-1.5$ (blue)}
	\label{fig:pp06}
\end{figure}
%
\begin{figure}
	\centering
		\includegraphics[scale=0.3,angle=90]{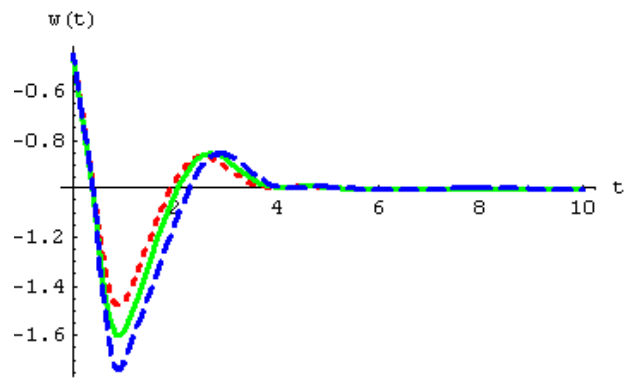}
	\caption{Equation of state $w(t)$ obtained with $f^{(3)}(t_0)=-0.5$ (red), $-1$ (green), $-1.5$ (blue)}
	\label{fig:ww06}
\end{figure}
%
\begin{figure}
	\centering
		\includegraphics[scale=0.3,angle=90]{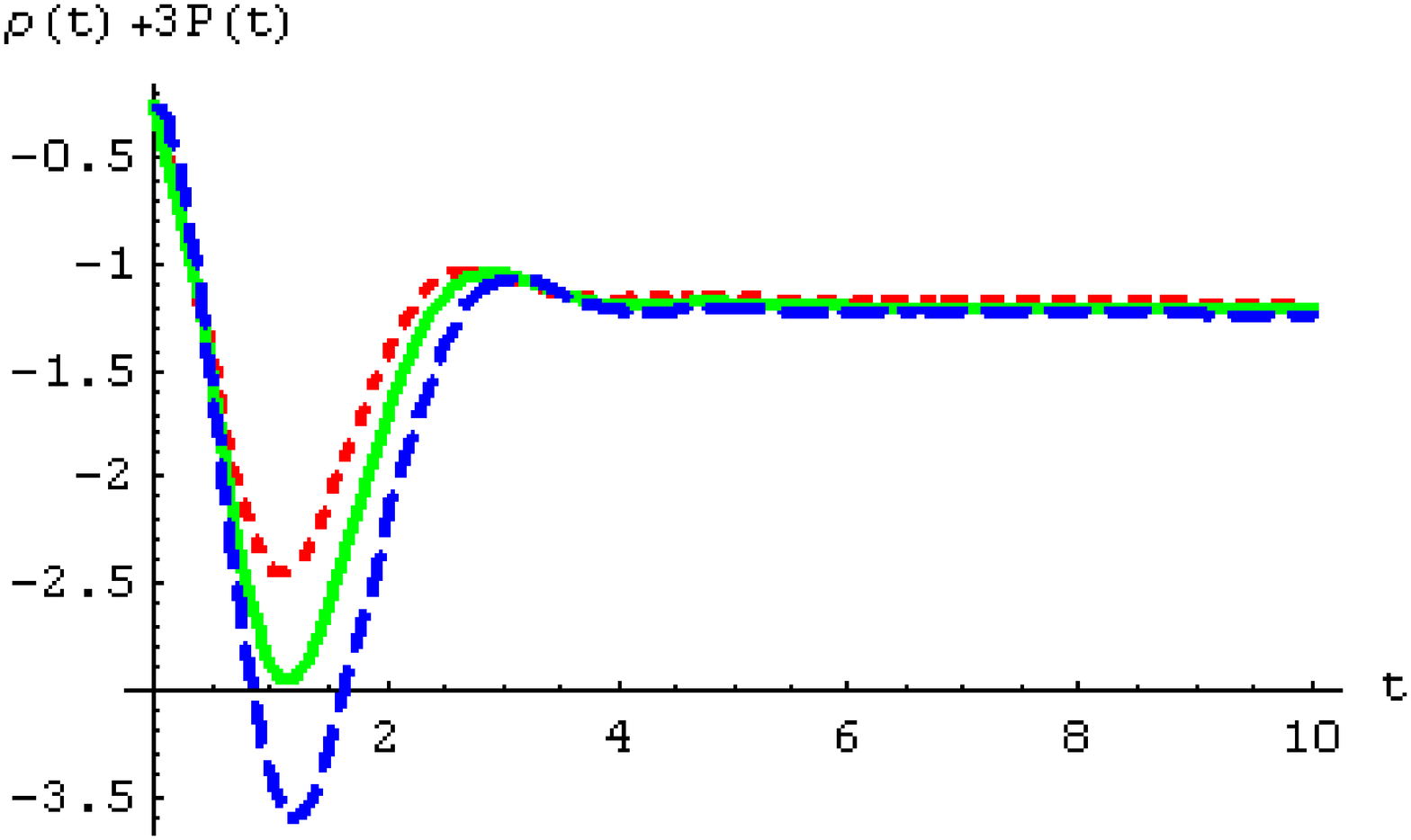}
	\caption{$\r(t)+3 p(t)$ obtained with $f^{(3)}(t_0)=-0.5$ (red), $-1$ (green), $-1.5$ (blue)}
	\label{fig:rp06}
\end{figure}

\end{center}
\pagebreak

\section{Conclusion}

From the plots obtained in the last section we find that the equation of
state $w\left( t\right) $ can take, at least, values between $-2$ and $0$.
Most interestingly it always converges to $-1$ after some time for all the
initial conditions we used. It means that the trace part of the diff field $%
D_{\mu \nu }$ acts like a cosmological constant $\Lambda $ in later times.
This scalar field spends most of the time before $w=-1$ as a phantom dark
energy. This is a time-dependent phantom dark energy and its effects on
Universe will be different from the time-independent cases [\ref{Ca1}, \ref%
{CW}, \ref{Gr}]. Since the equation of state $w$ can be $0$ in the very
early Universe, it behaves as a dark matter when it happens. The fact that
the value of $w$ never gets close to $1/3$ is very desirable because
otherwise the radiation-like behavior of the scalar field at early times
would influence already well explained primordial nucleosynthesis [\ref%
{BKLMS}].

The effects on the scalar field on the Hubble parameter are also notable. It
is easy to see that there are expected peaks [Eq.\ref{Hubs01}]. Figure \ref{fig:hh01}, Figure \ref{fig:hh03},
Figure \ref{fig:hh04}, and Figure \ref{fig:hh05} show shoulder-like regions where the value of
Hubble parameter stay almost constant, i.e., the universe is expanding
exponentially over this region:%
\begin{equation}
H=\frac{\dot{a}}{a}={\mbox{constant}}  \rightarrow a\left(
t\right) \sim e^{Ht}.
\end{equation}%
This scalar field, hence, acts like an inflaton over these regions. The
constant Hubble parameter in later time can be interpreted as the current
expansion of the Universe due to the cosmological constant like behavior of
the scalar field. The last important observation is that the value of $\rho
\left( t\right) +3p\left( t\right) $ is almost always negative. That is,
this scalar contributes to the acceleration of Universe for most of time.

Though our results are qualitative, they are showing very interesting
cosmological implications of the diffeomorphism field $D_{\mu \nu }$: the
trace part of the diffeomorphism field $D_{\mu \nu }$ can produce the
inflation-like behavior of the Universe in early time and the accelerating
Universe in later time qualitatively. It is also observed that the
time-dependent phantom dark energy plays a crucial role in this model. Our
results, however, lack the realistic prediction: $e$-foldings $N$,%
\begin{equation}
N\left( t\right) =\ln \frac{a\left( t_{\mbox{\tiny end}}\right) }{a\left( t\right) }%
=\int_{t}^{t_{end}}H\left( t\right) dt,
\end{equation}%
during ``inflation" predicted by this model is about $N\sim 2$, whereas any
realistic model for inflation requires $N\sim 70$. This suggest that more
research in the study the set of parameters and the initial values necessary
to make sensible predictions. Furthermore the understanding of the
time-dependent phantom dark energy is essential in future research.\vfill%
\pagebreak 

\section*{Acknowledgements}
VGJR would like to thank A.P. Balachandran, Y. Meurice, V.P. Nair, and P. Ramond for discussion. This work was supported by NSF grant PHY 02-44377.  

\end{document}